\documentclass[fleqn,10pt]{wlscirep}

\usepackage[utf8]{inputenc}
\usepackage[T1]{fontenc}
\usepackage[separate-uncertainty = true, multi-part-units = repeat]{siunitx}
\usepackage{amsmath,bm}
\usepackage{float}
\usepackage{comment}
\usepackage{multirow}
\usepackage{mathtools}

\newcommand{\TBADEN}{{$\mathrm{TBAD}_{EN}$~}} 
\newcommand{\TBADEX}{{$\mathrm{TBAD}_{EX}$~}}
\newcommand{\TBADOR}{{$\mathrm{TBAD}_{OR}$~}}
\newcommand{\DP}{{$\Delta P_{TL-FL}$}}
\newcommand{\DAOEN}{{$\mathrm{DAo}_{\mathrm{prox}}$~}}
\newcommand{\DAOEX}{{$\mathrm{DAo}_{\mathrm{dist}}$~}}
\newcommand{\FLFR}{\mathrm{FR}_{\mathrm{FL}}} 
\newcommand{\QFL}{\overline{\mathrm{Q}}_{\mathrm{FL}}}  
\newcommand{\QTL}{\overline{\mathrm{Q}}_{\mathrm{TL}}} 

\DeclareSIUnit\mmHg{mmHg}

\title{
Hemodynamic Effects of Entry and Exit Tear Size in Aortic Dissection Evaluated with \textit{In Vitro} Magnetic Resonance Imaging and Fluid-Structure Interaction Simulation
}

\author[1,2,$\ddag$]{Judith Zimmermann}
\author[1,$\ddag$,$\ast$]{Kathrin B\"aumler}
\author[1,5]{Michael Loecher}
\author[1,3,5]{Tyler E. Cork}
\author[3,4]{Alison L. Marsden}
\author[1,5]{Daniel B. Ennis}
\author[1]{Dominik Fleischmann}

\affil[1]{Stanford University, Department of Radiology, Stanford, CA}
\affil[2]{University of California, San Francisco, Department of Radiology and Biomedical Imaging, San Francisco, CA}
\affil[3]{Stanford University, Department of Bioengineering, Stanford, CA}
\affil[4]{Stanford University, Department of Pediatrics, Stanford, CA}
\affil[5]{Veterans Affairs Health Care System, Division of Radiology, Palo Alto, CA}

\affil[$\ddag$]{Contributed equally}
\affil[$\ast$]{Corresponding author, email: baeumler@stanford.edu}


\begin{abstract}

Understanding the complex interplay between morphologic and hemodynamic features in aortic dissection is critical for risk stratification and for the development of individualized therapy.
This work evaluates the effects of entry and exit tear size on the hemodynamics in type B aortic dissection by comparing fluid-structure interaction (FSI) simulations with \textit{in vitro} 4D-flow magnetic resonance imaging (MRI).
A baseline patient-specific 3D-printed model and two variants with modified tear size (smaller entry tear, smaller exit tear) were embedded into a flow- and pressure-controlled setup to perform MRI as well as 12-point catheter-based pressure measurements.
The same models defined the wall and fluid domains for FSI simulations, for which boundary conditions were matched with measured data.
Results showed exceptionally well matched complex flow patterns between 4D-flow MRI and FSI simulations.
Compared to the baseline model, false lumen flow volume decreased with either a smaller entry tear (\SIlist[list-units=single]{-17.8;-18.5}{\percent}, for FSI simulation and 4D-flow MRI, respectively) or smaller exit tear (\SIlist[list-units=single]{-16.0;-17.3}{\percent}).
True to false lumen pressure difference (initially \SIlist[list-units=single]{+11.0;+7.9}{\mmHg}, for FSI simulation and catheter-based pressure measurements, respectively) increased with a smaller entry tear (\SIlist[list-units=single]{+28.9;+14.6}{\mmHg}), and became negative with a smaller exit tear (\SIlist[list-units=single]{-20.6;-13.2}{\mmHg}).
This work establishes quantitative and qualitative effects of entry or exit tear size on hemodynamics in aortic dissection, with particularly notable impact observed on FL pressurization.
FSI simulations demonstrate acceptable qualitative and quantitative agreement with flow imaging, supporting its deployment in clinical studies.
\end{abstract}

\begin{document}
\flushbottom
\maketitle

\section*{Introduction}
Aortic dissection is a life-threatening cardiovascular disease, typically presenting with an acute and dramatic onset, followed by a chronic life-long phase of increased risk for late adverse events.
An estimate of up to \num{138000} individuals in the United States alone live with chronic aortic dissection~\cite{Fleischmann2022}.
Its pathological substrate and hallmark is the formation of a secondary flow channel within the aorta, the false lumen (FL), caused by abrupt delamination of the inner aortic wall layers.
The delaminated tissue that seperates the FL from the original true lumen (TL) is referred to as the dissection flap.
Blood flow enters the FL through the entry tear upstream, and returns to the TL downstream through one or more exit tears~\cite{Nienaber2016}. 
The outer boundary of the FL is mechanically weak and prone to fatal aortic rupture, particularly in the first 24 to 48 hours~\cite{Chiu2016}, which presents the most common cause of death in the acute phase.
Other major concerns are branch vessel ischemia resulting in organ malperfusion, and aneurysmal dilation of the FL.
For all patients reaching the chronic phase, life-long clinical monitoring is mandatory, and surgical or endovascular intervention is required in \SIrange[range-units=single]{40}{60}{\percent} of patients within the first five years~\cite{Pape2015,Afifi2015}.

Anatomical characteristics of aortic dissections vary substantially between patients.
Specifically, an aortic dissection without the involvement of the ascending aorta (Stanford type B aortic dissection, TBAD) is managed medically for the majority of patients (\SIrange[range-units=single]{60}{70}{\percent})~\cite{Fleischmann2022}.
However, studies suggest that late adverse events occur in \SIrange[range-units=single]{34}{38}{\percent} of initially uncomplicated TBADs~\cite{Fleischmann2022}.
According to latest guidelines~\cite{Isselbacher2022}, preventive thoracic endovascular aortic repair (TEVAR) may be considered for TBAD; nevertheless, the role and timing of TEVAR remains an evolving matter of debate.

Surveillance imaging is critical to identify potentially vulnerable patients and initiate intervention informed by measurable predictive biomarkers.
A number of previous works investigated morphological features that were mostly derived from computed tomography angiography (CTA) image data~\cite{Schwartz2018,Spinelli2018,Kunishige2006,Tsai2007,Sailer2017,Lavingia2015,Evangelista2012,Tolenaar2013a}.
Adding to morphological measures, recent studies increasingly investigated hemodynamics to further advance prediction models. For example, decreased outflow through FL branch vessels~\cite{Sailer2017}, increased FL ejection fraction~\cite{Burris2020,Marlevi2021,Zadrazil2020}, and FL pressurization~\cite{Tse2011,Cheng2014,Xu2020} were determined to promote late adverse events.
Further, these hemodynamic features have been associated with patient-specific morphology.
Cuellar-Calabria et al.~\cite{Cuellar-Calabria2021} presented entry tear dominance (i.e. entry tear much larger than exit tear) as a multivariable predictor, while Fleischmann and Burris~\cite{Fleischmann2021} hypothesize a linkage between tear dominance, increased outflow resistance, and resulting FL pressurization. That is, to optimize risk stratification, predictive models must comprise both morphological and hemodynamic indices.

Hemodynamic parameters can be obtained via computational fluid dynamics (CFD) and through flow imaging.
CFD simulates patient-specific flow fields and pressures at high spatio-temporal resolution, and fluid-structure interaction (FSI) additionally incorporates the effects of the deformable vessel wall~\cite{Baeumler2020, Baeumler2022}.
If simulations were able to reliably replicate \emph{in vivo} hemodynamics, it would tremendously expand non-invasive means to predict risk related to pathological changes and response to interventions.
Numerous studies have deployed CFD frameworks (with and without FSI) to study hemodynamics in patient-specific TBAD geometries~\cite{Karmonik2011,Shang2015,Cheng2014,Tse2011,Osswald2017,Xu2020}.
But, all face three key technical challenges:
1) it is difficult to obtain accurate patient-specific material parameters such as wall stiffness and fluid viscosity;
2) sparse clinical data, particularly of luminal pressure, often prevents a complete prescription of the boundary conditions;
and 3) lacking \emph{in vivo} measured flow and pressure data hinders rigorous validation.

4D-flow magnetic resonance imaging (MRI) measures time-resolved 3D vector-valued velocity maps within a 3D volume of interest~\cite{Markl2003a}, and has shown promising potential for measuring flow dynamics in dissections~\cite{DeBeaufort2019,Dillon-Murphy2016,Ruiz-Munoz2022}. Owing to lengthy scan times, however, clinical \textit{in vivo} 4D-flow MRI must be performed using high acceleration rates and ultra-fast sequences, but trades-off in spatio-temporal image resolution and signal-to-noise-ratio (SNR) are inevetiable ~\cite{Valvano2017,Ma2019,Schnell2014,Garg2018,Loecher2019,Dillinger2020}. Sub-optimal data quality ultimately hampers the retrieval of quantitative metrics, which in turn challenges the applicability of \textit{in vivo} 4D-flow MRI for validating simulations~\cite{Baeumler2020,Pirola2019}.

\begin{figure}[ht]
	\centering
	\includegraphics[]{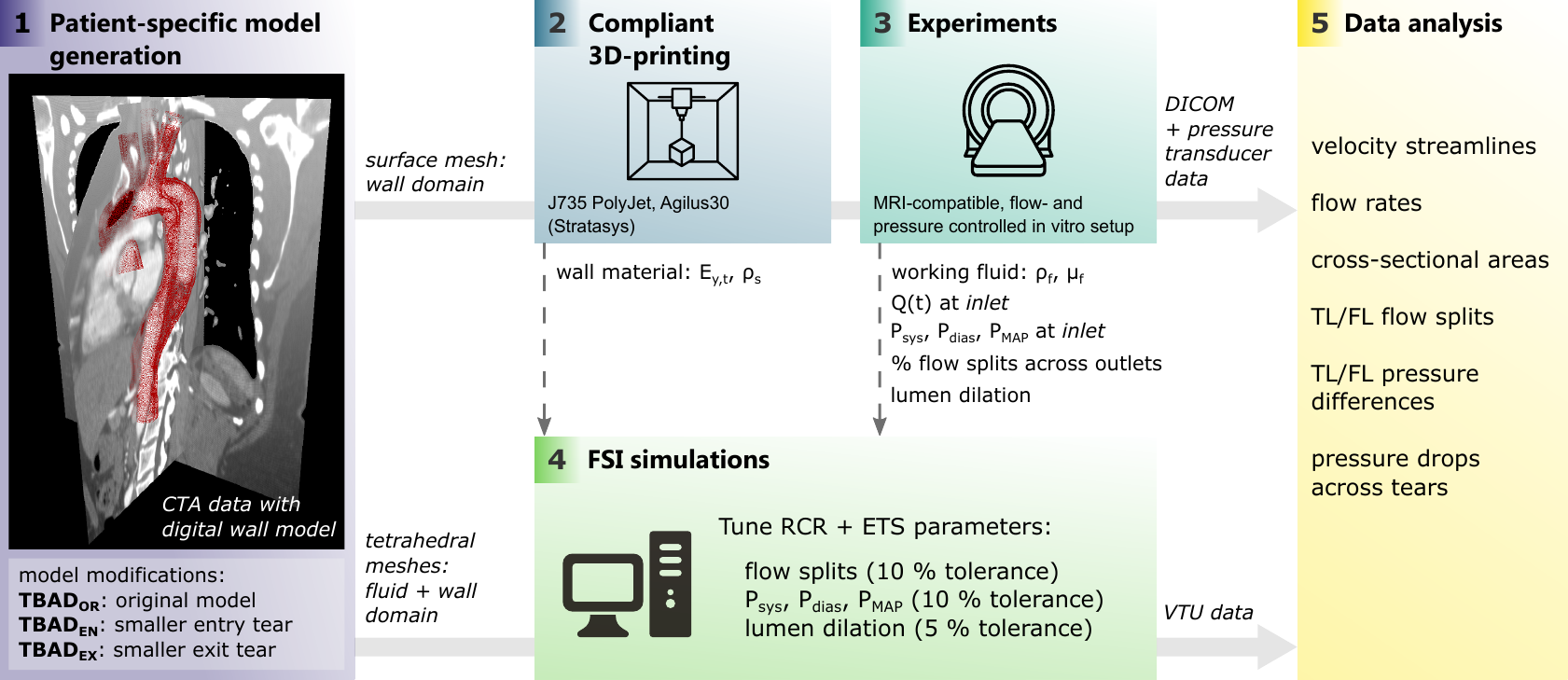}
	\caption{
		Pipeline showing the key methodological steps:
		(1) image-based patient-specific model generation from CTA data;
		(2) compliant 3D-printing;
		(3) experiments (MRI + catheter-based pressure mapping) with \textit{in vitro} setup;
		(4) FSI simulations with boundary conditions informed by measured data (dashed lines); and
		(5) data analysis, i.e. quantitative and qualitative comparison of hemodynamics between measured and simulated data.
		CTA: computed tomography angiogram,
		$Q(t)$: 2D-PC measured net flow,
		$P_{sys}$: systolic pressure,
		$P_{dias}$: diastolic pressure,
		$P_{MAP}$: mean arterial pressure,
		$E_{y,t}$: tangent Young's modulus wall,
		$\rho_s$: density wall,
		$\rho_f$: density fluid,
		$\mu_f$: dynamic viscosity fluid,
		RCR: Windkessel components,
		ETS: external tissue support,
		TL: true lumen,
		FL: false lumen.
		}
	\label{FIG_pipeline}
\end{figure}

We recently reported a flow- and pressure-controlled MRI-compatible setup which embeds subject-specific 3D-printed aortas with compliant walls that is appropriate for comparison with FSI simulations~\cite{Zimmermann2021c,Lan2022}.
The setup enables direct measurement of all boundary conditions and material parameters, and yields prolonged acquisition times for high-quality 4D-flow MRI data, thus overcoming the technical challenges of studies that deploy a CFD-only approach.
Herein, we build upon this work and generate \textit{in vitro} 4D-flow MRI from three TBAD geometries to establish a comprehensive comparison with FSI simulations (Fig. \ref{FIG_pipeline}).
The objectives of this study were:
(1) to evaluate the impact on hemodynamic metrics in TBAD amid variations in entry and exit tear area; and
(2) to compare these hemodynamic metrics as obtained from FSI simulations with \textit{in vitro} 4D-flow MRI.

\section*{Methods}
\subsection*{Patient-specific model generation}
\subsubsection*{Anatomy and imaging data}
A 3D CTA dataset of a 25 year old woman with an uncomplicated TBAD was selected retrospectively from our institutional database. CTA and all other methods were performed in accordance with standard of care procedures.
Data retrieval was approved by Stanford University Institutional Review Board (\#39377, ``Image Registry For Computer Simulations and Image-Based Modeling in Congenital and Acquired Cardiovascular Disease'', PI: Alison L. Marsden).
The requirement for written consent was waived due to the retrospective nature of the study.
Contrast enhanced CTA images were acquired at \SI{1}{\milli\meter} section thickness and \SI[product-units=single]{0.7x0.7}{\square\milli\meter} pixel size from the thoracic inlet to the common femoral arteries.
The imaging findings revealed an acute TBAD in the descending thoracic aorta with a dissection flap extending from the left subclavian artery origin down to the distal thoracic aorta, ending above the level of the diaphragm. The proximal entry tear into the false lumen was located immediately distal to the left subclavian artery, and a distal exit tear was located above the diaphragm, superior to the celiac artery origin. 
The cross-sectional area of entry and exit tear was \SI{228}{\square\milli\meter} and  \SI{227}{\square\milli\meter}, respectively. No branch vessels except for small intercostal arteries branched off of the TL or FL.

\subsubsection*{Segmentation and digital wall model}
\begin{figure}[t]
	\centering
	\includegraphics[]{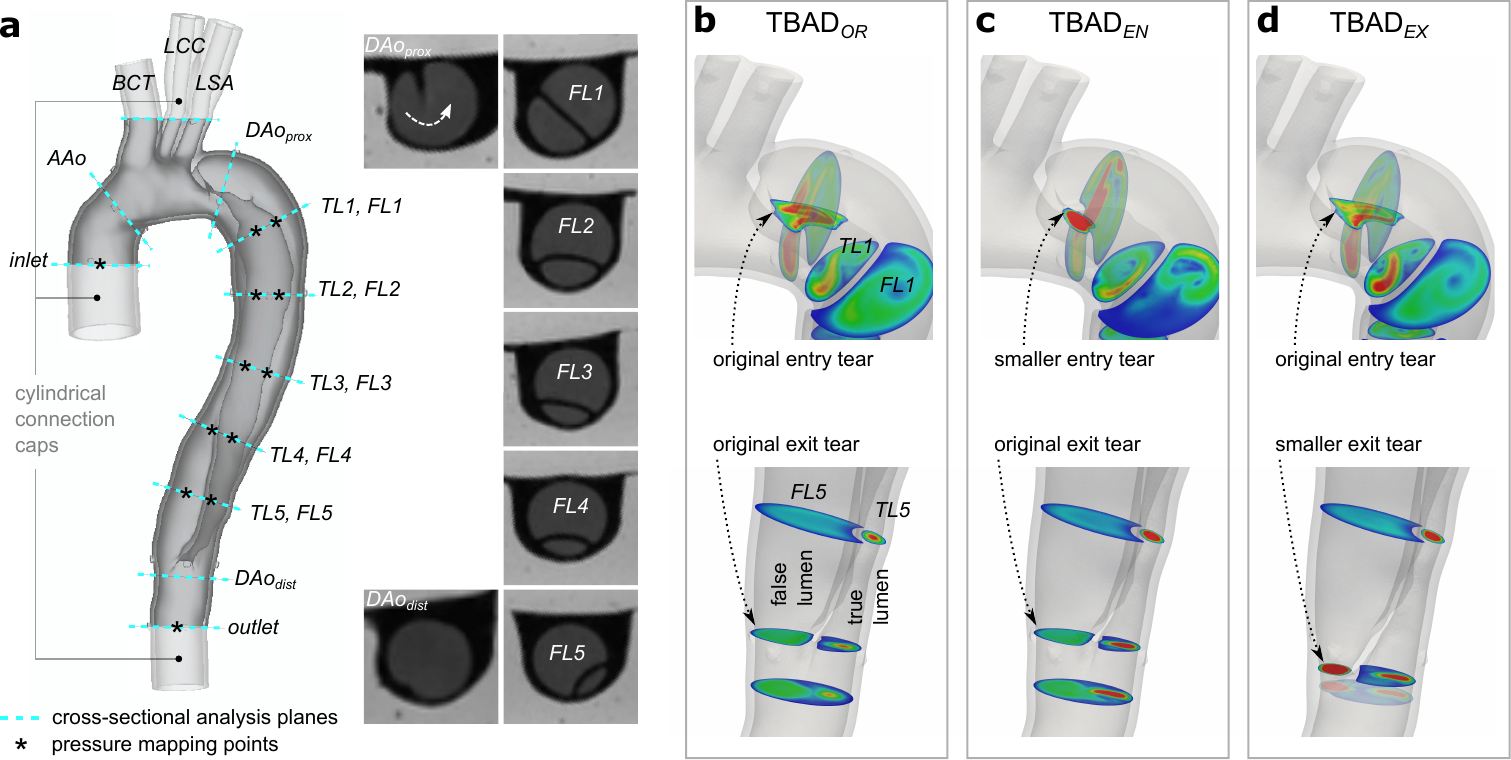}
	\caption{
		TBAD models.
		(a) Original digital wall model with definition of 18 landmarks that were defined for tuning and analysis purposes. Note that ${DAo}_{prox}$ cuts through the entry tear, while ${DAo}_{dist}$ is positioned just below the exit tear. Inset images show 2D-cine MRI frames for cross-sections along the dissected region.
		Close-up view of entry and exit tear regions for the (b) original model \TBADOR\!, (c) smaller entry tear model \TBADEN\!, and
		(d) smaller exit tear model \TBADEX\!.
		}
	\label{FIG_modelDescription}
\end{figure}

A digital model of the patient's thoracic aorta was generated from the CTA dataset. The model included the ascending aorta from above the sinotubular junction, the aortic arch, and the descending thoracic aorta to the level of the diaphragm. The three aortic arch branches were included in the model: brachiocephalic trunk (\textit{BCT}), left common carotid artery (\textit{LCC}), and left subclavian artery (\textit{LSA}). The intercostal arteries were excluded. The model domain was extended to \SI{35}{\milli\meter} distal to the re-entry tear.  
Two mesh domains were generated: 
The `fluid domain' -- representing the aortic lumen -- was segmented using active contours with manual refinements (itk-SNAP v3.4.0). 
The `wall domain' -- representing the outer aortic wall and the dissection flap -- was extruded from the fluid domain (Autodesk Meshmixer v3.5) with uniform thickness $h=\SI{2}{\milli\meter}$ (Fig. \ref{FIG_modelDescription}a). 
A detailed description of the TBAD model generation is provided in B\"aumler et al.~\cite{Baeumler2020}.
 
\subsubsection*{Digital Modification of Entry and Exit Tear Size}
The original patient-specific model \TBADOR was digitally modified to generate two additional models with reduced entry or exit tear size: \TBADEN with reduced entry tear and original exit tear size,
and \TBADEX with unchanged entry tear but reduced exit tear size (Fig. \ref{FIG_modelDescription}b, c, d).
The smaller size tears were $\approx\SI{27}{\percent}$ of their original size, measuring \SI{62}{\square\milli\meter} (entry tear) and \SI{61}{\square\milli\meter} (exit tear), respectively.
All models were augmented with visual landmarks to enable a precise definition of cross-sectional planes for data analysis and comparison. Additionally, the 3D-printed models were augmented with cylindrical caps to facilitate tubing connections.
Meshing and model modification tasks were performed using SimVascular~\cite{Updegrove2017} and Autodesk Meshmixer.

\subsection*{Compliant 3D-printing}
The wall models of \TBADOR\!, \TBADEN\!, \TBADEX were 3D-printed using a stereolithography technique (PolyJet J735, Stratasys Inc.) and a photopolymer print resin (Agilus30, Stratasys Inc.).
Printed models were finished with a thin coating film (Bectron, Elantas) to prevent fluid absorption and maintain the material characteristics during the flow experiments. 
The compliant print material underwent uniaxial tensile testing~\cite{Zimmermann2021c} and proved to be comparable to \textit{in vivo} aortic wall compliance (tangent Young's modulus $E_{y,t} = \SI{1.2}{\mega\pascal}$).

\subsection*{Experiments}
\subsubsection*{Flow setup}
An MRI-compatible flow setup (Supplementary Fig. S1) was assembled, driven by a programmable pump (CardioFlow 5000 MR, Shelley Medical Imaging Technologies) capable of providing highly-controlled flow and pressure conditions within a physiologic range~\cite{Zimmermann2021c}.
In brief, each model was connected to the circuitry tubing via custom designed 3D-printed barbed connectors with tapered transitions (Form 3B, Formlabs Inc.). Each model was then embedded into a solid ballistics gelatin block (ClearBallistics) for stabilization and to provide static MRI signal regions for velocity (phase) offset corrections. 
The model and gelatin mold were placed inside an enclosed box and connected through box-mounted flow fittings to the periphery. 
Two capacitance elements controlled the system's downstream compliance, and in turn, pressure waveform amplitude.
The net flow splits were adjusted through pinch valves distal to the capacitance elements; an ultrasonic flow sensor (ME-PXL14, Transonic) was used to fine tune the flow splits of the system.
The same pinch valves were used to control the outlet resistance and to tune the mean pressure conditions in the system. 
A pressure transducer (SPR-350S, Millar) was inserted through two ports to record pressure measurements at 12 locations along the model.

Glycerol-water (ratio = 40/60) with added MR contrast medium (ferumoxytol, concentration: \SI[per-mode=symbol]{0.75}{\milli\liter\per\liter}) was used as a blood-mimicking fluid with strong $T_1$ MRI signal enhancement. Fluid density ($\rho_f$) and dynamic viscosity ($\mu_f$) measurements confirmed $\rho_f=\SI{1.1}{\gram\per\cubic\milli\meter}$ and $\mu_f=\SI{0.0042}{\pascal\second}$. A typical aortic flow waveform was applied
(heart rate = \SI[per-mode=symbol]{60}{\per\minute}, 
stroke volume = \SI[per-mode=symbol]{74.1}{\milli\liter\per\second},
peak flow rate = \SI[per-mode=symbol]{300}{\milli\liter\per\second},
total flow = \SI[per-mode=symbol]{4.45}{\liter\per\minute}, Supplementary Fig. S1b).

\subsubsection*{System tuning and pressure mapping}
For each of the three models, pressure and flow split conditions were tuned on the MRI scanner table prior to image acquisition.
Tuning targets were defined as follows: 
flow split (outlet vs. combined arch branches) 70/30,
inlet systolic pressure ($P_s$) \SI{120}{\mmHg} or greater,
and inlet diastolic pressure ($P_d$) \SI{70}{\mmHg} or greater. 
The downstream resistance was reduced in the models with smaller entry or exit tear sizes (\TBADEN and \TBADEX compared to \TBADOR\!) to maintain pressure targets. 
Once tuning targets were met, luminal pressure data were recorded at twelve locations: one proximal to the dissection, five locations within the TL, five within the FL, and one distal to the dissection (Fig. \ref{FIG_modelDescription}a); three recordings of five consecutive pump cycles (five seconds) were obtained per landmark. 

\subsubsection*{MR imaging protocol}
MRI acquisitions were performed on a \SI{3}{\tesla} MRI scanner (Skyra, Siemens Healthineers) with a 32-channel spine and 18-channel body matrix coil. 
The pump provided an external trigger signal to retrospectively gate time-resolved sequences. Image data were acquired immediately after system tuning and pressure mapping. 

The protocol included sequences with parameters as follows:
(1) 3D spoiled gradient echo (3D-SPGR):
voxel size = \SI[product-units=single]{1.1x1.1x1.1}{\cubic\milli\meter}, 
FoV = \SI[product-units=single]{220x123}{\square\milli\meter},
TE = \SI{3}{\milli\second}, 
TR = \SI{5.25}{\milli\second},
flip angle = \ang{25}. 
3D-SPGR datasets were acquired under the following conditions: (i) no flow (``flow off'') and (ii) steady flow, thus no cardiac gating was applied. 
(2) 2D phase contrast (2D-PC) at eleven cross-sections (Fig. \ref{FIG_modelDescription}a):
pixel size = \SI[product-units=single]{1.1x1.1}{\square\milli\meter}, 
slice thickness = \SI{6}{\milli\meter}, 
FoV = \SI[product-units=single]{220x123}{\square\milli\meter},
TE = \SI{3}{\milli\second}, 
TR = \SI{5.25}{\milli\second}, 
flip angle = \ang{25}, 
averages = 2,  
$V_{enc}$ = 100--170 \si{\centi\meter\per\second},
retrospective gating (40 frames, temporal resolution = \SI{25}{\milli\second}).
(3) 2D cine gradient echo (2D-cine) at eleven cross-sections (Fig. \ref{FIG_modelDescription}a):
pixel size = \SI[product-units=single]{0.9x0.9}{\square\milli\meter}, 
slice thickness = \SI{6}{\milli\meter}, 
FoV = \SI[product-units=single]{240x150}{\square\milli\meter}
TE = \SI{3}{\milli\second}, 
TR = \SI{4.75}{\milli\second}, 
flip angle = \ang{7}, 
averages = 2, 
retrospective gating (40 frames, temporal resolution = \SI{25}{\milli\second}).
(4) 4D-flow with Cartesian k-space sampling:
FoV = \SI[product-units=single]{340x236x84}{\cubic\milli\meter}, 
matrix = \SI[product-units=single]{220x156x56}, 
voxel size = \SI[product-units=single]{1.5x1.5x1.5}{\cubic\milli\meter}, 
TE = \num{2.7} \si{\milli\second}, 
TR = \num{5.6} \si{\milli\second}, 
flip angle = \ang{15}, 
parallel imaging (GRAPPA, R=2), 
$V_{enc}$ = 120--180 \si{\centi\meter\per\second}, 
lines/segment = 2, 
retrospective gating (20 frames, temporal resolution = \SI{50}{\milli\second}).
Scan time per model was 1 hour and 30 minutes, and total end-to-end experiment time for all models was 12 hours.

\subsection*{FSI simulations}
This work's simulation approach was built on previously presented work on FSI simulations in a patient-specific aortic dissection case~\cite{Baeumler2020}. 
To match the experimental conditions for the present study, the following adaptations were made: (i) we prescribed homogeneous material properties throughout the structural domain, i.e. no distinction between intimal flap and outer material wall properties; (ii) prestress of the structural domain was set to zero. 

\subsubsection*{Mesh generation}
For each geometric model (\TBADOR\!, \TBADEN\!, and \TBADEX\!) we created refined unstructured tetrahedral meshes with the TetGen mesh generator~\cite{Si2008} (embedded in SimVascular~\cite{Updegrove2017}). 
Based on previous simulations, a mesh size of $\SI{1.3}{\milli\meter}$ was found to be sufficiently small~\cite{Baeumler2020}. 
Each mesh consisted of approximately $\num{1.1e6}$ tetrahedral elements for the fluid domain and approximately $\num{0.5e6}$ tetrahedral elements for wall domain.

\subsubsection*{Governing equations} 
The fluid flow was governed by the Navier–Stokes equations with viscosity $\mu_f = \SI{0.0042}{\Pa\second}$ and density $\varrho_f = \SI{1100} {\kg\per\cubic\meter}$. The FSI equations were formulated in arbitrary Lagrangian-Eulerian (ALE) coordinates~\cite{Bazilevs2010, Baeumler2020}, which allowed us to capture the coupled motion of the fluid and structural domain using a finite element method (described below). The glycerol-water mixture used in the experimental setup was modeled as incompressible and Newtonian.  
The 3D printed material representing the arterial wall was modeled as a homogeneous, isotropic, nonlinear, hyperelastic material, and described by a Neo-Hookean material model~\cite{Bazilevs2010}.
Structural density (per manufacturer's spec sheet) and elasticity (as measured) were $\varrho_s = \SI{1450} {\kg\per\cubic\meter}$ and $E_{y,t} = \SI{1.2}{\mega\pascal}$, respectively.
To account for the presence of the gelatin mold surrounding the printed models we applied external tissue support (ETS) to the outer arterial wall in the form of a Robin boundary condition~\cite{Moireau2012}:
\begin{equation}
   \boldsymbol{\sigma_S n} = -k_S \boldsymbol{u_S} - c_S \partial_t \boldsymbol{u_S} - p_0 \boldsymbol{n}, 
\end{equation}
where $\boldsymbol{\sigma_S}$ denotes the Cauchy stress tensor of the structural domain, $\boldsymbol{n}$ the unit outer normal vector, $\boldsymbol{u_S}$ the displacement field, and $p_0$ the external (or intrathoracic) pressure. The user-defined scalar parameters $k_S$ and $c_S$ denote the elastic and viscoelastic response of the external tissue, respectively.

\subsubsection*{Boundary data and tuning}
At the model inlet we prescribed a pulsatile flow waveform that was informed from \textit{in vitro} 2D-PC MRI measurements; a parabolic velocity profile was used.
At the four model outlets, we applied three-element Windkessel boundary conditions which account for downstream vascular effects~\cite{Vignon-Clementel2006}.
The total resistance $R_T$ and capacitance $C_T$ were tuned and distributed across all outlets such that every outlet $i$ was described by a proximal resistance $R_{P, i}$, distal resistance $R_{D, i}$, and capacitance $C_i$. The ratio of distal to proximal resistance $k_d$ was fixed for each model, and the distribution of total resistance and capacitance were governed by the experimentally determined flow splits. Details of the tuning process have been described in B\"aumler et al.~\cite{Baeumler2020}. 
The target pressure values comprised systolic pressure ($P_{sys}$), diastolic pressure ($P_{dias}$), and mean pressure ($P_{MAP}$) values --- informed by experimental pressure mapping. Target flow ratios across outlets \textit{BCT}, \textit{LCC}, \textit{LSA}, and \textit{outlet} were informed by 2D-PC net flow measurements.
Tuning targets (Table \ref{Table:TuningResults}) were matched with relative residual errors of $\SIlist{\leq1.4;\leq4.1;\leq4.6}{\percent}$ for pressure and $\SIlist{\leq3.4;\leq7.4;\leq7.3}{\percent}$ for flow splits
(corresponding to \TBADOR\!, \TBADEN\!, and \TBADEX\!, respectively)
\begin{table}[ht]
\footnotesize
\begin{center}
\begin{tabular}{clllllllll}
\multicolumn{1}{l}{}         &                                          & \multicolumn{3}{c}{pressure tuning at \textit{inlet}}            &                                          & \multicolumn{4}{c}{flow split tuning across outlets} \\[1ex]
\multicolumn{1}{l}{}         &                                          & $P_{sys}$ & $P_{dias}$ & $P_{MAP}$                      &                                          & \textit{BCT}     & \textit{LCC}     & \textit{LSA}     & \textit{outlet}  \\[1ex]
\multicolumn{1}{c|}{\TBADOR}  & \multicolumn{1}{l|}{simulated (mmHg)}    & 127.5  & 59.1    & \multicolumn{1}{l|}{86.1}   & \multicolumn{1}{l|}{simulated (\%)}   & 15.5   & 3.2     & 5.6     & 75.7   \\
\multicolumn{1}{c|}{}        & \multicolumn{1}{l|}{measured (mmHg)}     & 126.4  & 60.0    & \multicolumn{1}{l|}{86.1}   & \multicolumn{1}{l|}{measured (\%)}    & 15.0    & 3.1     & 5.4   & 76.5   \\
\multicolumn{1}{c|}{}        & \multicolumn{1}{l|}{residual error (\%)} & 0.82 & 1.42  & \multicolumn{1}{l|}{0.08} & \multicolumn{1}{l|}{residual error (\%)} & 3.37  & 3.22  & 3.08  & 1.02  \\ \hline\\[0.5ex]

\multicolumn{1}{c|}{\TBADEN} & \multicolumn{1}{l|}{simulated (mmHg)}    & 141.9  & 72.7    & \multicolumn{1}{l|}{95.7}   & \multicolumn{1}{l|}{simulated (\%)}   & 12.3   & 3.0   & 6.1   & 78.6   \\
\multicolumn{1}{c|}{}        & \multicolumn{1}{l|}{measured (mmHg)}     & 137.7  & 69.9    & \multicolumn{1}{l|}{96.2}   & \multicolumn{1}{l|}{measured (\%)}    & 11.4   & 2.8   & 5.7   & 80.1   \\
\multicolumn{1}{c|}{}        & \multicolumn{1}{l|}{residual error (\%)} & 3.02 & 4.11  & \multicolumn{1}{l|}{0.57} & \multicolumn{1}{l|}{residual error (\%)} & 7.43  & 7.21  & 7.00  & 1.84  \\ \hline\\[0.5ex]

\multicolumn{1}{c|}{\TBADEX} & \multicolumn{1}{l|}{simulated (mmHg)}    & 148.6  & 75.1    & \multicolumn{1}{l|}{100.4}  & \multicolumn{1}{l|}{simulated (\%)}   & 11.7   & 3.3   & 5.0    & 80.1   \\
\multicolumn{1}{c|}{}        & \multicolumn{1}{l|}{measured (mmHg)}     & 142.1  & 73.5    & \multicolumn{1}{l|}{98.9}   & \multicolumn{1}{l|}{measured (\%)}    & 10.9   & 3.1   & 4.6   & 81.4   \\
\multicolumn{1}{c|}{}        & \multicolumn{1}{l|}{residual error (\%)} & 4.58   & 2.24    & \multicolumn{1}{l|}{1.5}    & \multicolumn{1}{l|}{residual error (\%)} & 7.26  & 7.06  & 6.93  & 1.63 \\ \hline
\end{tabular}
\caption{
		Inlet pressure (\si{\mmHg}) and flow splits ($\si{\percent}$ of total outflow) for the three models. The FSI simulations matched target pressure values (systolic ($P_{sys}$), diastolic ($P_{dias}$), and mean ($P_{MAP}$)) with a relative error of $\leq\SI{4.58}{\percent}$ and absolute error of $\leq\SI{6.5}{\mmHg}$. Flow split ratios between outlets were matched in the FSI simulations with a relative error of $\leq\SI{7.43}{\percent}$ and an absolute error or $\leq\SI{1.5}{\percent}$.
		}
\label{Table:TuningResults}
\end{center}
\end{table}

ETS parameters $k_S$ and $c_S$ were chosen such that the FSI-simulated minimum-to-maximum lumen dilation at \textit{AAo} matched the MRI-measured minimum-to-maximum dilation in the 3D-printed models.
ETS tuning yielded a relative dilation of 
\SIlist{5.0;5.0;5.3}{\percent} in FSI compared to
\SIlist{4.4;5.0;5.4}{\percent} in 2D-cine MRI 
(for \TBADOR\!, \TBADEN\!, and \TBADEX\!, respectively).
Tuning parameter settings and pressure traces are provided in Supplementary Table S2 and Supplementary Fig. S3, respectively.

\subsubsection*{Numerical solver}
The governing equations were solved with the open source finite element solver svFSI (SimVascular~\cite{Updegrove2017,Zhu2022}). 
svFSI features a second order generalized $\alpha$-time stepping scheme, linear tetrahedral elements for pressure and velocity (with pressure and momentum stabilization), two-way coupling of the fluid and structural domain, pre-conditioning of the resulting linear systems, and backflow stabilization at the fluid outlets~\cite{Marsden2015,Esmaily-Moghadam2015,Esmaily-Moghadam2011}. 
To assure periodicity of the simulation, at least 6 cardiac cycles were simulated before results were extracted for the final cycle.
The simulations were carried out on two nodes (Intel(R) Xeon(R) Gold 5118, 2.30 GHz) with a total of 48 CPUs; approximate runtime was ten hours per cardiac cycle.


\subsection*{Data analysis}
For MRI data, TBAD lumen segmentation was performed on high-resolution 3D-SPGR steady flow data (using watershed-based region-growing) and facilitated image-based flow visualization.
2D-cine datasets were used to automatically track through-cycle aortic wall and flap deformation~\cite{Tautz2012} and to compute cross-sectional area change. 2D-PC datasets were corrected for velocity (phase) offsets and processed to retrieve the inlet flow rate waveform as well as percentage net flow splits across outlets.
4D-flow datasets were corrected for the following conditions: (i) Maxwell terms (during MR reconstruction), (ii) gradient non-linearity distortion~\cite{Markl2003}, and (iii) phase offsets. 4D-flow datasets were reformatted at each analysis location (Fig. \ref{FIG_modelDescription}a) to generate cross-sectional phase and magnitude images. Reformatted magnitude images were used to track lumen boundaries which served to calculate quantitative flow parameters based on reformatted phase images. 
Post-processing of the MRI data was completed using MevisLab modules (v3.5a, Fraunhofer Institute for Digital Medicine) and ParaView (v5.7, Kitware).
Pressure measurements were analyzed using dedicated data acquisition system software (LabChart 8, ADInstruments) and pressure signals were averaged over five cardiac cycles.
FSI simulation results were exported as VTK unstructured grid files (.vtu) from svFSI at 80 time-steps over a single cardiac cycle (i.e. temporal frame length: $\SI{12.5}{\milli\second}$). Qualitative visualization and quantitative analysis of simulated datasets were performed using ParaView (v.5.7, Kitware).

\section*{Results}

\subsection*{Flow patterns}
Each of the three models exhibited a unique flow pattern and velocity distributions, with excellent qualitative agreement between 4D-flow MRI and FSI simulations (Figs. \ref{FIG_tubingViz}, \ref{FIG_vecViz}, and Supplementary Videos S4-S8).
Highly complex and distinct flow patterns were observed in the proximal false lumen adjacent to the entry tear region. 
A high-velocity flow jet through the entry tear with impingement on the opposite FL wall was only observed in the \TBADEN model.
The impingement zone was also associated with a steep local pressure gradient affecting the FL wall (Fig. \ref{FIG_vecViz}c).
Helical flow patterns were also visible in the proximal TL close to the entry tear and were well captured by both techniques (Fig. \ref{FIG_tubingViz}b).
In the distal part of the dissected descending aorta, the modified models, \TBADEN and \TBADEX, showed increased flow velocities in the TL (Fig. \ref{FIG_tubingViz}a, arrows), which was consistent with the higher TL net flow volumes in these models, as reported below.
In addition, \TBADEX showed increased velocity through the exit tear compared to the \TBADOR and \TBADEN\! models. The smaller exit tear in the \TBADEX model also caused reversed flow distal to the exit tear (Fig. \ref{FIG_tubingViz}c, Supplementary Video S8), whereas \TBADOR and \TBADEN showed unidirectional laminar flow in this region.
During diastole, additional flow oscillations through the entry tear (in particular for \TBADEN\!, Fig. \ref{FIG_vecViz}b), and persistent helical flow in FL (Supplementary Videos S6, S7) were observed.

\subsection*{Flowrate measurements}
\subsubsection*{FL flow ratio}
The FL flow ratio ($\FLFR$, in \si{\percent}) was defined based on the TL and FL net flow volumes ($\QTL$, $\QFL$) as:

\begin{equation}
    \FLFR \coloneqq \frac{\QFL}{\QFL + \QTL}*100.
\end{equation}

\noindent $\FLFR$ was substantially decreased in the modified tear models, \TBADEN and \TBADEX, compared to the original \TBADOR\! model.
Specifically, $\FLFR$ was
\SIlist{73.7;55.2;56.4}{\percent}
for 4D-flow MRI and
\SIlist{77.6;59.8;61.6}{\percent}
for FSI simulations (values given for \TBADOR\!, \TBADEN\!, and \TBADEX\!, respectively).

\begin{figure}[H]
	\centering
	\includegraphics[]{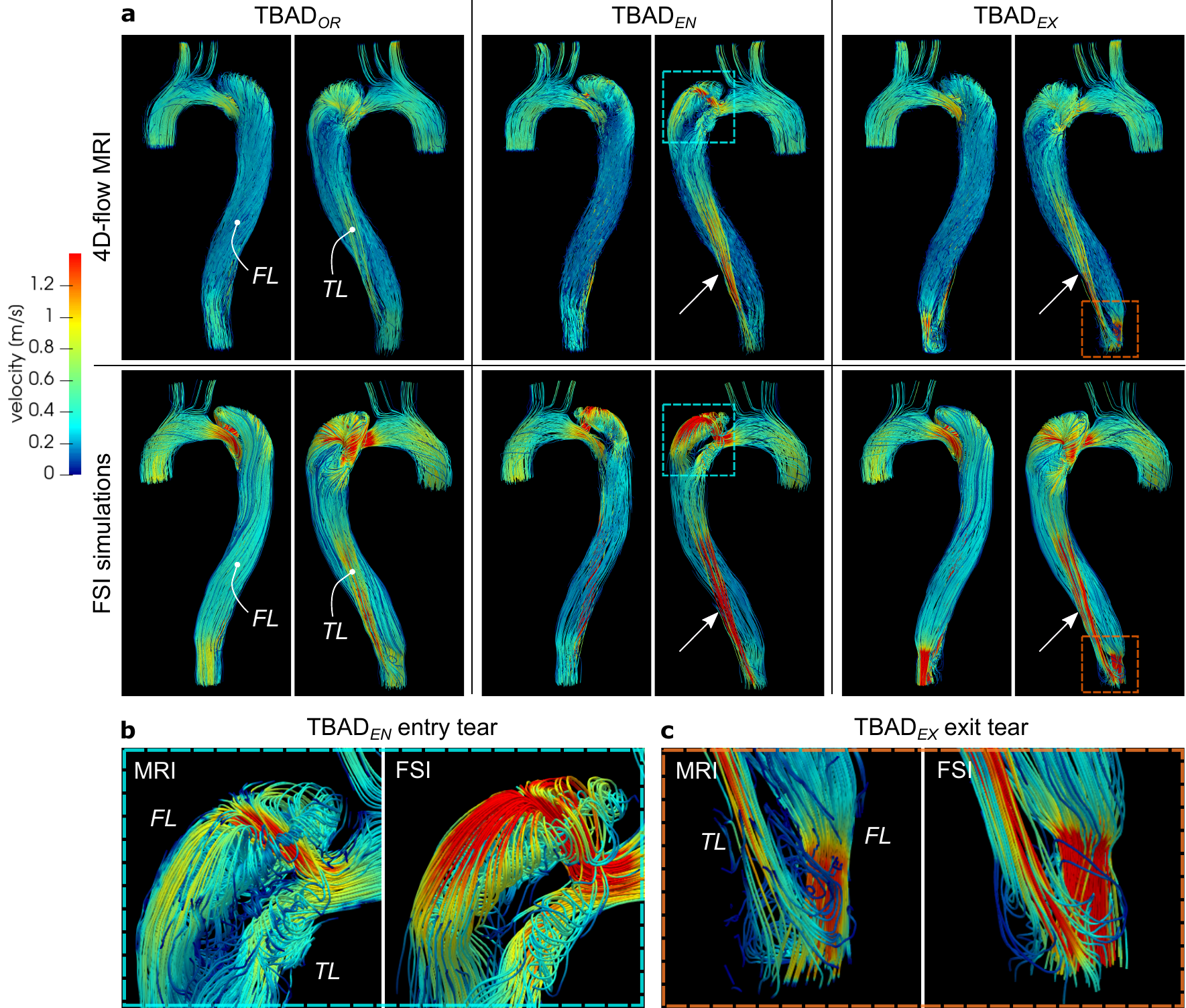}
	\caption{
		(a) Streamlines at peak systole ($t=\SI{200}{\milli\second}$) rendered from 4D-flow MRI (top) and FSI-simulated (bottom) data.
		Each model exhibits unique local flow characteristics that are in agreement between techniques. Key observations include:
		(i) increased flow velocities through the entry tear region, particularly in 
	    \TBADEN\!, and local helical flow in the proximal TL and FL in the vicinity of the entry tear (blue box, close-up view in (b));
		(ii) increased TL flow velocity for the modified \TBADEN and \TBADEX\! models (arrows);
		(iii) flow jet through small size exit tear in \TBADEX, with recirculating TL flow distal to the exit tear (orange box, close-up view in (c)). 
        Graphics created using ParaView (v5.7, \url{https://www.paraview.org}).
		}
	\label{FIG_tubingViz}
\end{figure}

\begin{figure}[H]
	\centering
	\includegraphics[]{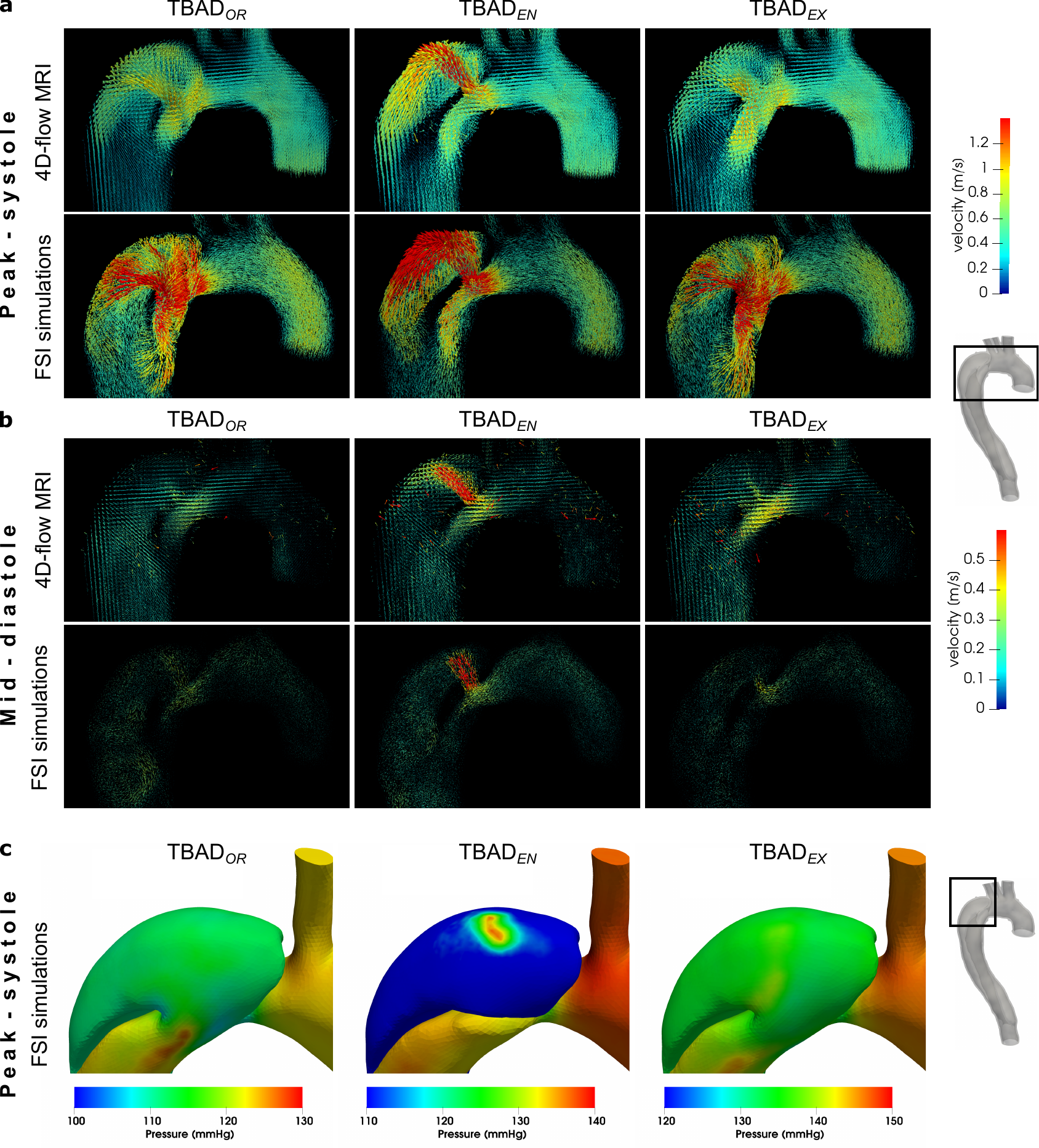}
	\caption{
		Flow and pressure dynamics in the arch region. 
		Velocity vectors in the aortic arch at (a) peak systole ($t=\SI{200}{\milli\second}$) (b) and mid-diastole  ($t=\SI{600}{\milli\second}$) based on 4D-flow MRI data and FSI simulations.
		During systole, flow patterns are in agreement between both modalities and for each model, but velocities are higher in FSI simulations. 
		Mid-diastole renderings reveal a secondary push through the entry tear, that is most pronounced in \TBADEN\!. For full-cycle animations refer to Supplementary videos 1-5. 
		(c) Absolute pressure at the lumen boundary ($t=\SI{250}{\milli\second}$) from the FSI simulations. \TBADEN shows a local pressure difference of $\SI{\approx35}{\mmHg}$ in the FL impingement zone. Graphic created using ParaView (v5.7, \url{https://www.paraview.org}).}
	\label{FIG_vecViz}
\end{figure}

\subsubsection*{Dynamic flowrates over the cardiac cycle}
\begin{figure}[H]
	\centering
	\includegraphics[]{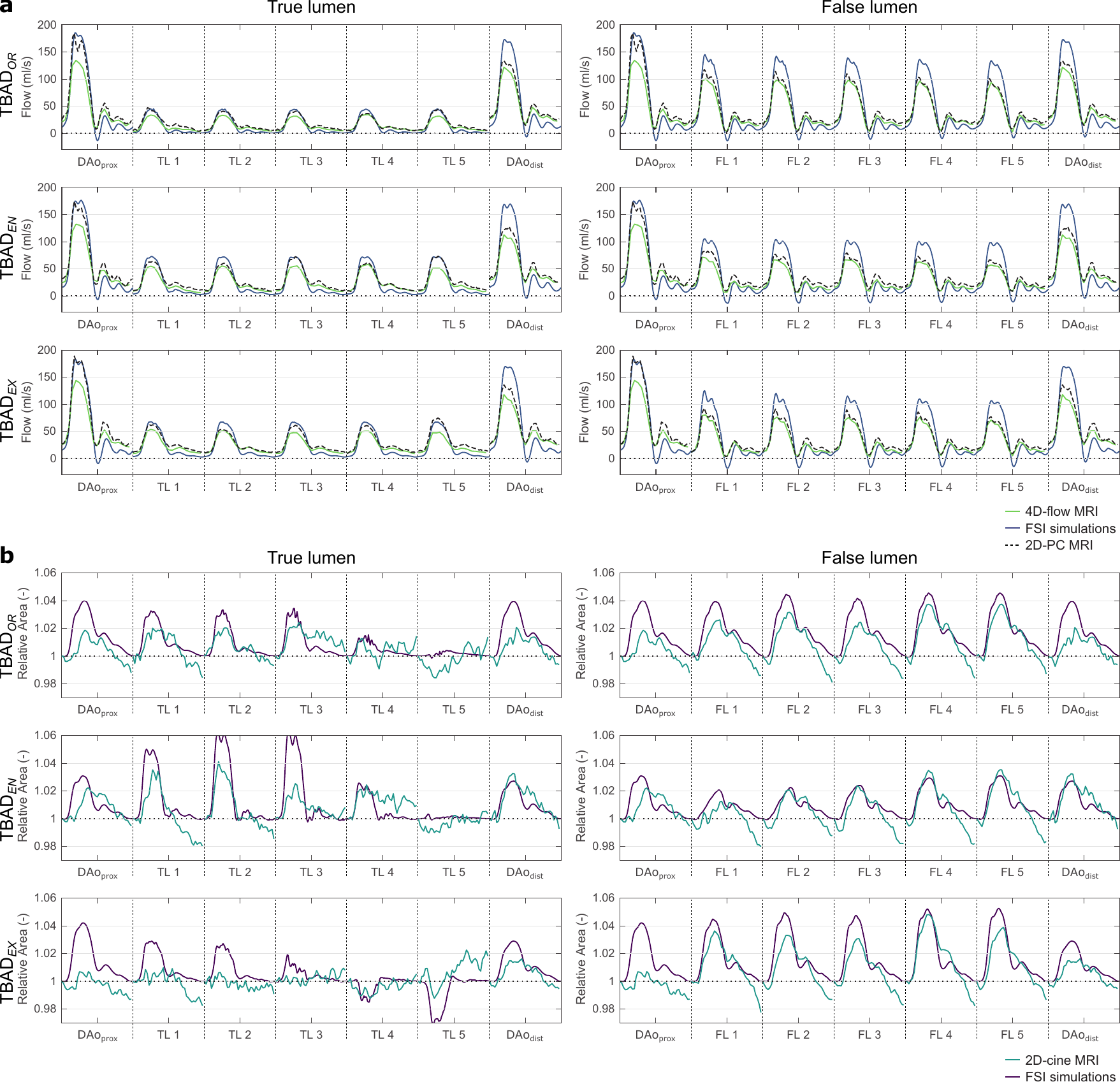}
	\caption{
		(a) Flow rates over the cardiac cycle along the dissected descending aorta based on 4D-flow MRI (green), FSI simulations (blue), and 2D-PC MRI (black, dashed). 
		(b) Cross-sectional area over the cardiac cycle based on 2D-cine MRI (light blue) and FSI simulations (purple).
		Relative area change was defined as absolute area divided by area in the first cycle frame.
		For landmark label definition see Fig. \ref{FIG_modelDescription}a.}
	\label{FIG_transient_flow_area}
\end{figure}

Transient flow rates at multiple cross-sectional landmarks showed higher systolic peaks in FSI simulations, whereas higher diastolic flow was observed in 4D-flow MRI (Fig. \ref{FIG_transient_flow_area}a). 
Damping of the flow waveform from \DAOEN to \DAOEX as well as along the individual lumina was observed in all three models, with stronger damping in 4D-flow and 2D-PC data than in FSI simulations.
Diastolic flow oscillations were most pronounced in FSI simulations, with short periods of reverse (i.e. negative) flow at $t \approx \SI{500}{\milli\second}$.
To further assess potential underestimation of systolic flow in 4D-flow MRI, 2D-PC measurements at identical locations were added to Fig. \ref{FIG_transient_flow_area}a.

\subsection*{Cross-sectional area measurements}
The cross-sectional lumen area was evaluated as absolute area, 
and as relative area changes over the cardiac cycle.
For the latter, the first  frame was used as the end-diastolic reference for normalization.
In addition to area measurements during pulsatile flow, we also obtained absolute area values from the STL model file, from both ``flow-off'' and steady flow MRI imaging datasets, and from steady flow FSI simulations. These additional measurements were all based on the \TBADOR\! model.

\subsubsection*{Absolute area measurements}

FSI simulations, compared to 2D-cine MRI based wall contour tracking, resulted in smaller absolute cross-sectional area at the majority of landmarks (Fig. \ref{FIG_pressureDrop_area_v2}b).
Specifically, FSI-based area measurements in peak-systole and end-diastole (mean~$\pm$~SD) differed from MRI-based values by
$\SIlist{
-16.6 \pm 10.5;
-7.0 \pm 11.2;
-3.1 \pm 12.2}
{\percent}$
for \TBADOR\!, \TBADEN\!, and \TBADEX\! respectively.
Additional ``flow-off'' experiments showed acceptable agreement with STL-based values ($\SI{-4.7\pm16.1}{\percent}$), 
but measurements during steady flow affirmed smaller absolute cross-sectional areas of the simulated deformable wall domain than in the printed model ($\SI{-0.6\pm10.4}{\percent}$).
Numerical results for all area measurements are listed in Supplementary Table S9 and plotted in Supplementary Fig. 10.

\subsubsection*{Relative area changes over cardiac cycle}
Figure \ref{FIG_transient_flow_area}b displays transient curves of relative cross-sectional areas for \TBADOR\!, \TBADEN\!, and \TBADEX\!, respectively:
TL areas increased by a factor of up to
\numlist{1.02;1.04;1.03}
in the experiments and up to
\numlist{1.03;1.06;1.03}
in FSI simulations over the cardiac cycle.
FL areas increased by a factor of up to
\numlist{1.04;1.04;1.05}
in experiments and up to
\numlist{1.05;1.03;1.05}
in FSI simulations
 for \TBADOR\!, \TBADEN\!, and \TBADEX\!, respectively.
TL in \TBADEX showed a slight collapse at the distal landmarks \textit{TL4} ($\approx \SI{-1}{\percent}$) and \textit{TL5} ($\approx \SI{-3}{\percent}$), as captured best in FSI simulations;
2D-cine MRI data did not fully reveal this behavior.


\subsection*{Pressure}

\subsubsection*{Pressure drops along centerline}
\begin{figure}[H]
	\centering
	\includegraphics[]{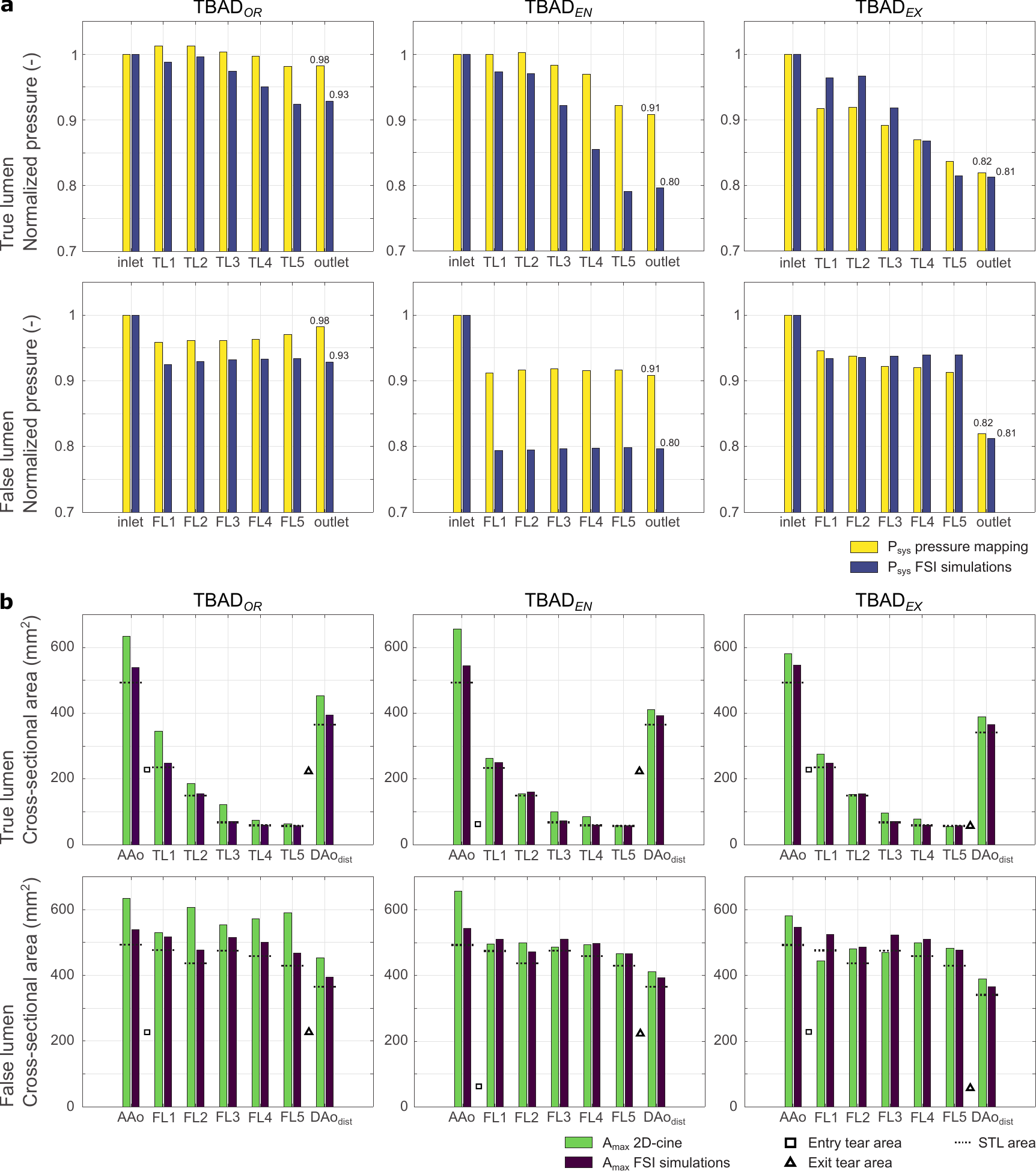}
	\caption{
		(a) Relative pressure (i.e. normalized to peak pressure at \textit{inlet}) drops along the aortic centerline including \textit{inlet}), \textit{outlet}), as well five landmarks in both the TL and FL.
		(b) Maximum cross-sectional area measurements at the identical landmarks used for pressure mapping, except for \textit{AAo}, which was chosen over 'inlet', since \textit{inlet} is fixed both in the 3D-printed model and in the simulation setup. Dashed bars denote area values obtained from the STL model file, squares denote size of entry tear, and triangles denote size of exit tear. For landmark label definition see Fig. \ref{FIG_modelDescription}a.}
	\label{FIG_pressureDrop_area_v2}
\end{figure}

Peak-systolic pressure ($P_{sys}$) at twelve pressure mapping landmarks was normalized with respect to the peak-systolic pressure at \textit{inlet} (Fig. \ref{FIG_pressureDrop_area_v2}a).
Overall, \textit{inlet} to \textit{outlet} pressure drops were smallest in the original \TBADOR model ($\SIlist{2;7}{\percent}$),
and larger for the modified \TBADEN 
($\SIlist{9;20}{\percent}$)
and \TBADEX
($\SIlist{18;19}{\percent}$);
given value pairs correspond to experiments and FSI simulations.

Pressure drop across the entry tear (i.e. TL entering FL, see \textit{inlet} to \textit{FL1}) was largest in \TBADEN\!, with a decrease of 
$\SIlist{17;20}{\percent}$ 
in experiments and FSI simulations, respectively.
Pressure drop across the exit tear (i.e. FL merging back into the TL, see \textit{FL5} to \textit{outlet}) was largest in \TBADEX\!, with a pressure decrease by 
$\SIlist{11;14}{\percent}$
in experiments and FSI simulations, respectively.

For all models, peak-systolic pressure did not drop noticeably along the FL centerline (\textit{FL1} to \textit{FL5}). However, pressure values steadily decreased along the TL (\textit{TL2} to \textit{TL5}), with the strongest TL pressure decline observed in \TBADEX\! ($\SIlist{15;20}{\percent}$ in the experiments and FSI simulations, respectively). 
The largest discrepancy between experimental pressure measurements and FSI simulations was found in the \TBADEN\! model, particularly in the FL. 
Here, \textit{inlet} to FL pressure (at \textit{FL1}) dropped by
$\geq \SI{20}{\percent}$
in FSI simulations, but
$\leq \SI{10}{\percent}$
in the experiments.

\subsubsection*{TL-FL pressure differences}
\begin{figure}[H]
	\centering
	\includegraphics[]{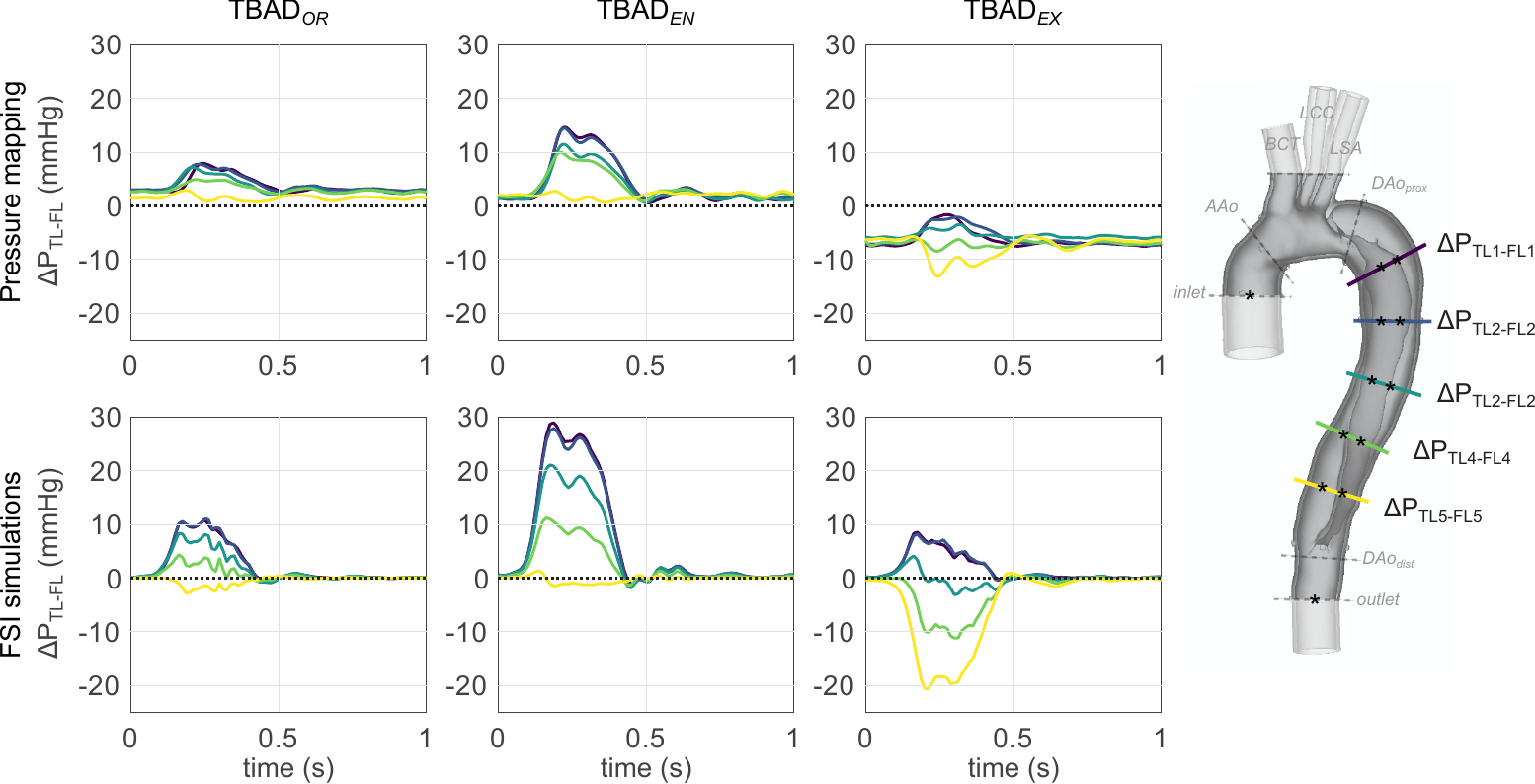}
	\caption{
		Inter-luminal pressure difference \DP\! between TL and FL according to pressure transducer measurements (top) and FSI simulations (bottom).
		\DP\! increased when the entry tear was made smaller (\TBADEN\!), and decreased --- including negative pressure difference at distal landmarks --- when exit tear made smaller (\TBADEX\!).}
	\label{FIG_pressureDiff_v2}
\end{figure}

Inter-luminal pressure differences were calculated as 
\begin{equation}
    \Delta P_{TL-FL} \coloneqq P_{TL} - P_{FL}
\end{equation}
at five locations along the dissected aorta.
\DP was substantially affected by tear size and steadily decreased from proximal to distal locations (Fig. \ref{FIG_pressureDiff_v2}).
\DP was positive at all landmarks and throughout the cardiac cycle for \TBADOR and \TBADEN\!, with maximum systolic differences of 
(\num{7.9}, \num{11.0})
\si{\mmHg}
for \TBADOR and 
(\num{14.6}, \num{28.9})
\si{\mmHg}
for \TBADEN\!.
\TBADEX FL pressure exceeded TL pressure by up to
(\num{13.2}, \num{20.6})
\si{\mmHg}
during systole; values are given for catheter-based and simulated data, respectively.

At end-diastole, FSI-simulated TL and FL pressures were in approximate equilibrium 
($\SI{-0.5}{\mmHg}<\Delta P_{TL-FL}<\SI{1}{\mmHg}$), whereas
experimental pressure measurements resulted in $\Delta P_{TL-FL}\neq0$. Inter-luminal diastolic pressure differences measured slightly positive for the \TBADOR and \TBADEN models
($\SI{0.9}{\mmHg}<\Delta P_{TL-FL}<\SI{3.2}{\mmHg}$), and negative for the \TBADEX model
($\SI{-7.6}{\mmHg}<\Delta P_{TL-FL}<\SI{-5.5}{\mmHg}$).

\section*{Discussion}
Hemodynamic factors --- together with biomechanical, mechanobiological, genetic, and morphological factors --- underlie the development and progression of aortic dissection and are being investigated as biomarkers for prognosis and treatment decisions.
The complex interplay between these factors is incompletely understood, and evolving computational methods to obtain patient-specific hemodynamic must be rigorously validated.
The present study assessed the hemodynamic effects of entry and exit tear size variations in a patient-specific TBAD model using two modalities.
We directly compared hemodynamic features of FSI simulations against \textit{in vitro} MRI and catheter-based pressure data in compliant 3D-printed TBAD models. 
The \textit{in vitro} approach enabled the study of quantitative hemodynamics in a highly controlled environment and without a scan time limitation.
Moreover, the experiments informed the boundary conditions and material parameter specifications in the FSI simulations with catheter-based pressure data and high fidelity material property estimates, which are not typically available in any clinical setting~\cite{Baeumler2020}.

Local and global flow helices, recirculation zones, and flow jets through tears were visualized in fine detail via streamlines,  and were well captured in both 4D-flow MRI and FSI simulations.
In particular, local helices in the vicinity of the entry tear (Fig. \ref{FIG_tubingViz}b, \TBADEN\!) and distal to the exit tear (Fig. \ref{FIG_tubingViz}c, \TBADEX\!) are strikingly similar using both techniques.
The emergent entry tear flow jet in \TBADEN impinging on the opposite FL wall, creating a local pressure gradient was equally demonstrated by 4D-flow MRI and FSI (Fig. \ref{FIG_vecViz}a).
This impingement zone may promote tissue degradation and destructive remodeling via mechanobiological pathways and potentially lead to aneurysmal degeneration~\cite{Meng2014,Wang2022}.

Comparing results between both modalities (FSI and MRI), in systole, FSI simulations displayed overall higher flow velocities, whereas in diastole, 4D-flow measured higher velocities.
Similarly, simulated flow rates were higher in systole and lower in diastole when compared to MRI data, specifically in 4D-flow MRI (Fig. \ref{FIG_transient_flow_area}a).
Two confounding effects may explain this result:
First, there is general consensus that 4D-flow MRI underestimates peak velocities when compared to 2D-PC MRI. Our results confirm that 2D-PC MRI consistently resulted in higher peak velocities, and therefore showed smaller discrepancies with FSI simulations.
Second, less dampening seems to occur in FSI simulations.
This is nicely reflected by the initially well matched flow rates (between 2D-PC and FSI) at landmark \textit{inlet}, but then higher systolic and lower diastolic flow (for FSI simulations) at all downstream landmarks.
 
Entry and exit tear size considerably affected inter-luminal pressure differences and true and false lumen flow splits.
Peak-systolic \DP\! was positive in the \TBADOR model (up to \SI{11.0}{\mmHg}), then further increased in the reduced entry tear model (up to \SI{28.9}{\mmHg}), but flipped to negative in the reduced exit tear model (up to \SI{-20.6}{\mmHg}).
Additionally, both of the tear-modified models significantly reduced $\FLFR$ down to \SI{75}{\percent} of the initial \TBADOR value.
This drastic $\FLFR$ reduction indicates that the flow throughput is dictated by the total resistance and independent of the location of the narrowing.
Consequently, it appears that despite the decrease in $\FLFR$ an increase in outflow resistance --- which only occurred in the \TBADEX model --- contributes to FL pressurization, which is thought promote FL degeneration and aneurysm formation.

Inter-modality comparison showed overall acceptable agreement for \DP.
Both FSI simulations and catheter-based measurements link a small exit tear to false lumen pressurization, which is hypothesized to increase the risk for late adverse events in TBAD patients. 
Moreover, our results align with findings by Cuellar-Calabria et al.~\cite{Cuellar-Calabria2021}, who identified ``entry tear dominance'' (here: \TBADEX model) as a predictor of late adverse events.
Absolute numbers for FSI-derived \DP\! deviated from catheter-based measurements and showed
(1) greater systolic inter-luminal pressure gradients, and
(2) diastolic inter-luminal pressure equilibrium at all landmarks, rather than \DP$\neq0$ in catheter-based measurements.
To explain these effects, we revisited potential causes as follows: differences of actual tear size between the 3D-printed model and the digital wall model, wall and fluid mesh coarseness distal to the exit tear, and catheter-based measurement inaccuracies relative to measured \DP.
But, none of these additional analyses led to a sound explanation as to why these discrepancies occured.

Tear size alterations also affected peak-systolic pressure drops along the luminal centerline.
First, simulated and measured data agreed on substantially greater \textit{inlet} to \textit{outlet} pressure drops in both of the tear-modified models.
Similar trends that describe increased pressure drops have been reported for aortic coarctation~\cite{Riesenkampff2014,Urbina2016}, which refers to a focal narrowing of the proximal descending aorta.
Second, the modification of tear size led to different locations of the steepest pressure drop along the centerline (Fig. \ref{FIG_pressureDrop_area_v2}).
\TBADEN exhibited the largest gradient across the entry tear; whereas \TBADEX exhibited the largest gradient across the exit tear.
Although simulated data presented larger pressure drops between centerline locations, relative trends again were well-matched between modalities.

In addition to the effect of entry and exit tear sizes, our results also showcase the relationship between the cross-sectional luminal area and the $P_{sys}$ drops along the respective luminal centerlines.
Along the TL centerline, $P_{sys}$ decreased incrementally, while the cross-sectional area narrowed considerably further downstream --- that is, the area at \textit{TL5} was \SI{24}{\percent} of the area at \textit{TL1}.
In comparison, our results suggest no considerable drop of $P_{sys}$ along the FL centerline, along which cross-sectional area values remained approximately constant.

The combined results for \DP\! and relative area change over the cardiac cycle also revealed the well known clinical phenomenon of TL compression and collapse.
If only FSI simulation are considered, one can deduce:
\DP$>0$ caused an increase in TL area,
\DP$\approx0$ was reflected by negligible area change (that is, constant area through the cycle), and \DP$<0$ (as observed in the \TBADEX model only) caused a decrease in TL area, corresponding to a compression of the TL.
Area measurements from 2D-cine MRI also hint at potential TL collapse in the \TBADEX\! model.

Three limitations of the present study should be addressed.
First, several assumptions simplified the actual TBAD \textit{in vivo} scenario:
Both experiments and simulations utilized a uniform thickness and elasticity for the outer aortic wall and the dissection flap. Thus, we are unable to analyze effects of spatially varying or non-isotropic wall characteristics on hemodynamics.
The FSI simulations also did not pre-stress the wall domain. While this matches the \textit{in vitro} measurements for a 3D-printed models, any study that considers inherently pre-stressed \textit{in vivo} tissue and data may require incorporation pre-stress effects~\cite{Baeumler2020}.
In addition, the ETS modeling parameters in the FSI simulations were applied uniformly. Both the presented \textit{in vitro} embedding of the models, and likely also the surrounding of the thoracic aorta \textit{in vivo}, is expected to non-uniformly restrict the movement of the outer aortic wall.

Second, retrieving precise wall deformation measurements of the 3D-printed models was limited by 2D-cine MRI spatial resolution as well as the inherent error of the registration-based wall tracking algorithm.
Specifically, this affected area measurements in the relatively small-sized TL.
While obtaining structural motion from FSI simulations can be considered error-free, measuring area based on the given 2D-cine MRI is not.
Future works must refine methods for measuring wall and flap motion in the experimental setup.

Lastly, this study investigated only a single TBAD case with its unique patient-specific features; 
specifically, no fenestration points (that is, additional small communications between the TL and FL along dissection flap) were present, the dissection did not extend distal to the celiac trunk, and --- for feasibility purposes --- we excluded intercoastal arteries when building the model.
Novel advances in 3D printing, specifically the integration of tissue mimicking materials, provide excellent versatility in model manufacturing regarding global and local geometry, thickness and elasticity --- a technical tool that could be leveraged in future works.

In conclusion, this work describes changes in the TBAD hemodynamics due to tear size alterations while comparing results from 
\textit{in vitro} MRI and pressure mapping experiments with FSI simulations.
In particular, the results demonstrate FL pressurization owing to a decreased exit tear size --- with well-matched observations between measurements and simulation.
The present study contributes to a better understanding of the interplay between TBAD morphology and associated quantitative hemodynamics.

\section*{Acknowledgements}
We thank Shannon Walters, Chris LeCastillo, and Kyle Gifford (Stanford 3DQ Lab) for their 3D printing support and services, and Nicole Schiavone (Stanford Mechanical Engineering) for advice about the technical setup.
We acknowledge the computational infrastructure (Sherlock HPC cluster) provided by the Stanford Research Computing Center.
Funding was received through DAAD doctoral candidate scholarship (to J.Z.), NIH R01 LM013120 (to A.L.M.) and NIH R01 HL131823 (to D.B.E.) 

\section*{Author contributions}
J.Z. and K.B. jointly conceptualized the study, performed cross-modality data analysis, prepared figures, and wrote the manuscript.
J.Z. engineered the flow loop setup and led experimental MRI studies. 
K.B. developed the computational simulations framework and performed FSI simulations.
M.L. and T.E.C. supported experimental data acquisition and reconstruction.
A.L.M. conceptualized the study and supervised computational simulation work.
D.B.E. conceptualized the study and supervised experimental data acquisition.
D.F. conceptualized the study and provided overall advice to the research objective.
All authors critically revised and approved the manuscript.

\section*{Additional Information}
\textbf{Data availability:}
TBAD model files (.stl), MRI data (.dcm), pressure traces (.csv), and FSI-simulated data (.vtu) will be made publicly available upon publication.\newline

\bibliography{library}
\newpage
\section*{Supplementary Material}
See below for supplementary figures and tables S1-S3, S9, S10. 
Upon publication we will make supplementary videos (.mp4) available. These are numbered S4-S8, with captions printed at the end of this document.

\vspace{1cm}

\begin{figure}[H]
    \centering
    \includegraphics[]{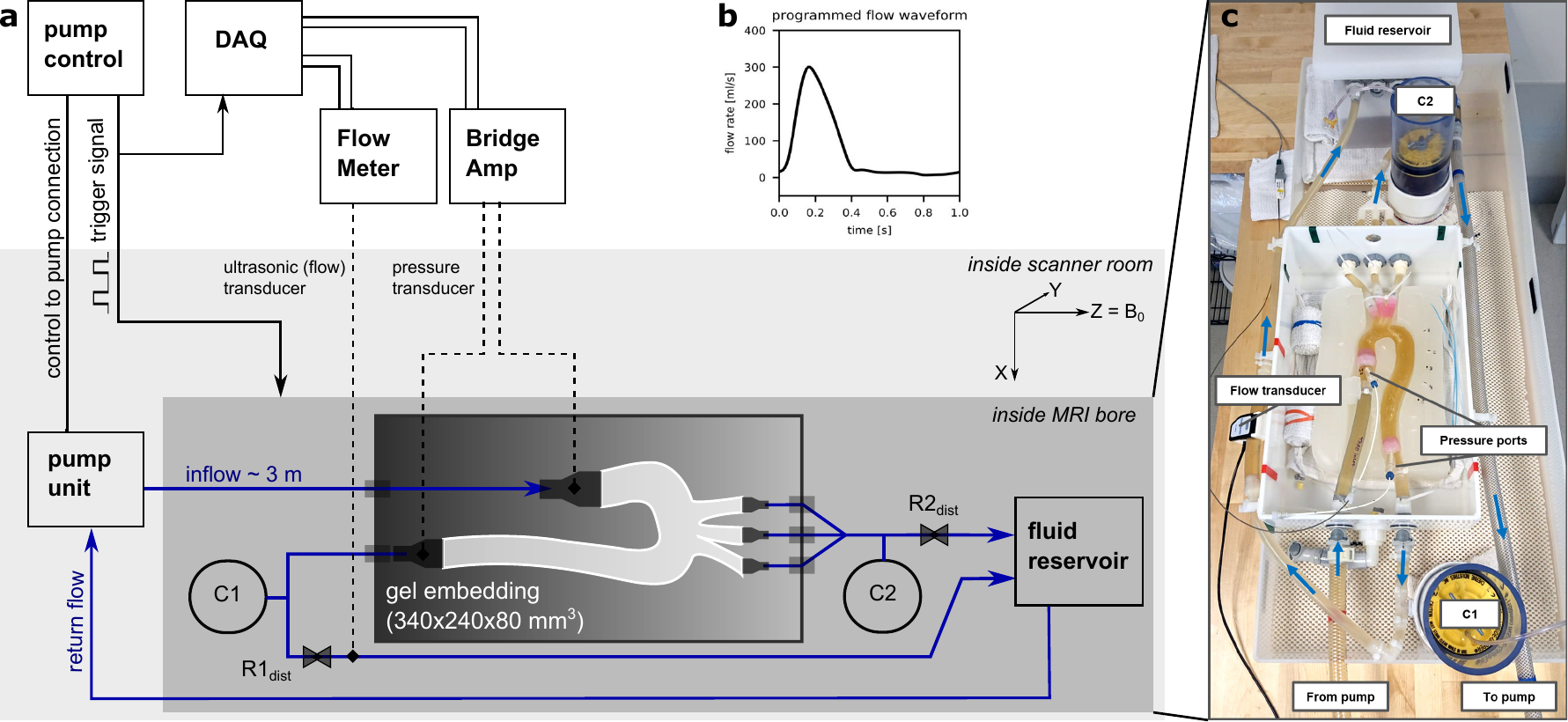}
    \caption*{
 \textbf{S1 Experimental setup.}
 (a) Schematic drawing of the flow loop setup; fluid lines are displayed in blue.
 (b) Programmed flow rate waveform with stroke volume of \SI[per-mode=symbol]{74.1}{\milli\liter\per\second}.
 (c) Photograph of the setup (with gel top and inner box lid both removed). All shown parts were inside the scanner bore during image acquisition. All other equipment (not shown in photograph) was positioned either outside the bore (pump unit), or outside the scanner room (pump control, data acquisition system (DAQ), bridge amplifier, flow meter).}
\end{figure}

\vspace{1cm}
\begin{table}[H]
\footnotesize
\begin{center}
\begin{tabular}{ c|| p{1.6cm}|  p{1.6cm}|   p{0.7cm}||  p{1.5cm}| p{1.5cm}|| p{1.2cm}| p{1.2cm}| p{1.2cm}| p{1.2cm}}
        & $R_T$
        & $C_T$  
        &  $k_d$ 
        & $k_s$  
        & $c_s$
        & $\varrho_f$
        & $\mu_f$
        & $\varrho_s$
        & $E_{y,t}$\\
        & (\si{\mega\Pa\second\per\cubic\meter})
        & (\si{\cubic\meter\per\Pa})
        & (-)
        & (\si{\mega\newton\per\cubic\meter})
        & (\si{\kilo\newton\second\per\cubic\meter})
        & (\si{\kg\per\cubic\meter})
        & (\si{\Pa\second})
        & (\si{\kg\per\cubic\meter})
        & (\si{\mega\Pa})\\[0.5ex]\hline
\TBADOR & 150 & \num{6.74 e-9}&     0.84& \num{-18}  &  \num{-30}
        & 1100 & 0.0042 & 1450 & 1.2 \\
\TBADEN & 161 & \num{1.23 e-8}& 0.87 & \num{-18} &  \num{-30} 
        & 1100 & 0.0042 & 1450 & 1.2 \\
\TBADEX & 170&  \num{1.02 e-8}&     0.86&  \num{-18} &  \num{-30}
        & 1100 & 0.0042 & 1450 & 1.2 \\
\end{tabular}
\caption*{ 
        \textbf{S2. FSI simulations parameters.}
        The three-element Windkessel boundary parameters include the total resistance $R_T$, total capacitance $C_T$, and ratio of distal to proximal resistance $k_d$.
        $R_T$ and $C_T$ are distributed across outlets according to measured flow splits (see Table 2 in main article) and the respective value for $k_d$.
        ETS scalar parameters $k_s$ (elastic response) and $c_s$ (viscoelastic response) were chosen to match the minimum-to-maximum dilation of the simulation to MRI-measured values.
        Fluid and structural density ($\varrho_f$ and $\varrho_s$), fluid viscosity $\mu_f$, and elastic modulus $E_{y,t}$ were prescribed according to benchtop measurements or manufacturer's information.}
\label{Table:TuningParameters}
\end{center}
\end{table}

\newpage
\begin{figure}[t]
    \centering
    \includegraphics[]{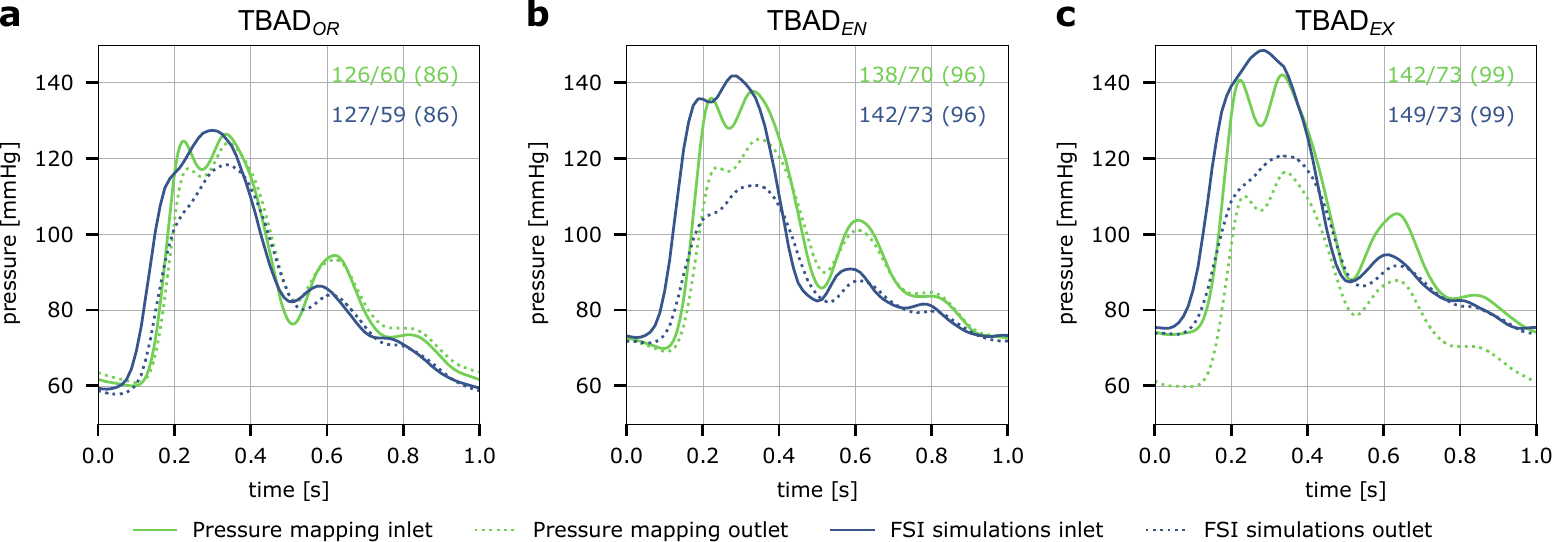}
    \caption*{
        \textbf{S3. Pressure waveforms} at inlet (solid) and outlets (dotted) for catheter-based measurements (green) and FSI simulations (blue): (a) \TBADOR\!, (b) \TBADEN\!, and (c) \TBADEX\!. Numbers in plotting areas report $P_{sys}$/$P_{dias}$ ($P_{\textrm{MAP}}$) in \si{\mmHg}.
        }
    \label{FIG_pressureTuning}
\end{figure}

\begin{table}[ht]
\footnotesize
\begin{center}
\begin{tabular}{ c  p{0.7cm}||  p{0.7cm}|   p{0.7cm}|   p{0.7cm}|   p{0.7cm}|   p{0.7cm}||  p{0.7cm}|   p{0.7cm}|   p{0.7cm}|   p{0.7cm}|   p{0.7cm}||  p{0.7cm}}
& \multicolumn{1}{c}{\textit{AAo}}  
& \multicolumn{1}{c}{\textit{TL1}}  
& \multicolumn{1}{c}{\textit{TL2}}  
& \multicolumn{1}{c}{\textit{TL3}}  
& \multicolumn{1}{c}{\textit{TL4}}  
& \multicolumn{1}{c}{\textit{TL5}}
& \multicolumn{1}{c}{\textit{FL1}}
& \multicolumn{1}{c}{\textit{FL2}}  
& \multicolumn{1}{c}{\textit{FL3}}
& \multicolumn{1}{c}{\textit{FL4}}
& \multicolumn{1}{c}{\textit{FL5}}  
& \multicolumn{1}{c}{$DAo_{dist}$}\\[0.8ex]
\multicolumn{8}{l}{\bf MRI (pulsatile flow, retrieved from 2D-cine data)}
\\[0.8ex]
\multicolumn{5}{l}{\TBADOR} \\
\multicolumn{1}{r|}{$A_0$}
    &$  612.4   $
    &$  339.1   $&$ 182.0   $&$ 118.8$&$    73.5    $&$ 62.3    $
    &$  517.4   $&$ 587.7   $&$ 540.3$&$    551.5   $&$ 569.5   $
    &$  444.7   $
    \\
\multicolumn{1}{r|}{$A_{max}$}  &$  633.9   $&
    $   345.9   $&$ 185.7   $&$ 121.7$&$    74.5    $&$ 63.2    $&
    $   530.9   $&$ 606.2   $&$ 553.5$&$    572.1   $&$ 590.9   $&
    $   453.9   $\\
\multicolumn{1}{r|}{$A_{mean}$} &$  620.2   $&
    $   340.9   $&$ 183.0   $&$ 120.3$&$    73.8    $&$ 62.2    $&
    $   521.3   $&$ 592.8   $&$ 543.4$&$    558.5   $&$ 577.0   $&
    $   447.2   $\\
\hline\\[0.5ex]
\multicolumn{5}{l}{\TBADEN}
\\[0.6ex]   
\multicolumn{1}{r|}{$A_0$}
    &$  630.3   $
    &$  253.7   $&$ 148.9   $&$ 97.3$&$ 83.7    $&$ 57.2    $
    &$  490.5   $&$ 489.6   $&$ 475.1$&$    477.0   $&$ 450.2   $
    &$  398.5   $
    \\
\multicolumn{1}{r|}{$A_{max}$}  &$  656.9   $&
    $   262.6   $&$ 155.0   $&$ 99.8$&$ 85.7    $&$ 57.5    $&
    $   496.6   $&$ 499.8   $&$ 486.4$&$    493.6   $&$ 466.1   $&
    $   411.5   $\\
\multicolumn{1}{r|}{$A_{mean}$} &$  640.1   $&
    $   254.4   $&$ 150.0   $&$ 98.0$&$ 84.8    $&$ 57.0    $&
    $   490.4   $&$ 492.0   $&$ 476.7$&$    480.8   $&$ 454.6   $&
    $   403.5   $\\
\hline\\[0.5ex]
\multicolumn{5}{l}{\TBADEX}
\\[0.6ex]   
\multicolumn{1}{r|}{$A_0$}  
    &$  559.0   $
    &$  272.2   $&$ 151.9   $&$ 95.0$&$ 77.6    $&$ 54.3    $
    &$  428.5   $&$ 466.2   $&$ 456.2$&$    477.1   $&$ 464.8   $
    &$  382.9   $
    \\
\multicolumn{1}{r|}{$A_{max}$}  &$  581.6   $&
    $   274.8   $&$ 152.9   $&$ 95.9$&$ 78.0    $&$ 55.5    $&
    $   444.1   $&$ 481.6   $&$ 470.3$&$    500.1   $&$ 482.9   $&
    $   389.1   $\\
\multicolumn{1}{r|}{$A_{mean}$} &$  566.8   $&
    $   271.7   $&$ 151.6   $&$ 95.3$&$ 77.3    $&$ 54.7    $&
    $   433.1   $&$ 470.3   $&$ 459.4$&$    484.8   $&$ 470.3   $&
    $   384.9   $\\
\hline\\[0.5ex]
    
\multicolumn{8}{l}{\bf MRI (steady flow, retrieved from 3D-SPGR data)}
\\[0.8ex]
\multicolumn{1}{r|}{\TBADOR}    
    &$  527.0   $
    &$  315.0   $&$ 154.0   $&$ 85.0$&$ 81.0    $&$ 65.0    $
    &$  547.0   $&$ 587.0   $&$ 553.0$&$    564.0   $&$ 457.0   $
    &$  457.0   $
    \\
\hline\\[0.5ex]
    
\multicolumn{8}{l}{\bf MRI (``flow-off'', retrieved from 3D-SPGR data)}
\\[0.8ex]
\multicolumn{1}{r|}{\TBADOR}    
    &$  435.0   $
    &$  268.0   $&$ 152.0   $&$ 85.0$&$ 83.0    $&$ 60.0    $
    &$  418.0   $&$ 431.0   $&$ 411.0$&$    438.0   $&$ 445.0   $
    &$  373.0   $
    \\
\hline\\[0.5ex]
    
\multicolumn{8}{l}{\bf FSI (pulsatile flow)}
\\[0.8ex]
\multicolumn{5}{l}{\TBADOR}
\\[0.6ex]
\multicolumn{1}{r|}{$A_0$}  
    &$  513.5   $
    &$  239.6   $&$ 150.7   $&$ 68.4$&$ 58.9    $&$ 56.9    $
    &$  497.2   $&$ 457.4   $&$ 494.8$&$    479.0   $&$ 448.0   $
    &$  379.6   $
    \\
\multicolumn{1}{r|}{$A_{max}$}  &$  538.7   $&
$   247.4   $&$ 155.7   $&$ 70.7$&$ 59.8    $&$ 57.1    $&
$   516.8   $&$ 477.7   $&$ 515.4$&$    500.8   $&$ 468.3   $&
$   394.6   $\\
\multicolumn{1}{r|}{$A_{mean}$} &$  523.2   $&
$   242.2   $&$ 152.1   $&$ 69.1$&$ 59.1    $&$ 56.9    $&
$   505.2   $&$ 465.5   $&$ 502.9$&$    487.5   $&$ 455.9   $&
$   385.5   $\\
\hline\\[0.5ex]
\multicolumn{5}{l}{\TBADEN}
\\[0.6ex]
\multicolumn{1}{r|}{$A_0$}  
    &$  518.5   $
    &$  238.3   $&$ 151.2   $&$ 68.7$&$ 59.0    $&$ 56.9    $
    &$  499.6   $&$ 462.2   $&$ 499.3$&$    483.7   $&$ 452.2   $
    &$  382.9   $
    \\
\multicolumn{1}{r|}{$A_{max}$}  &$  544.2   $&
$   250.2   $&$ 160.6   $&$ 73.0$&$ 60.5    $&$ 57.1    $&
$   509.9   $&$ 472.6   $&$ 511.2$&$    497.9   $&$ 466.2   $&
$   393.2   $\\
\multicolumn{1}{r|}{$A_{mean}$} &$  526.7   $&
$   241.5   $&$ 153.5   $&$ 69.8$&$ 59.4    $&$ 57.0    $&
$   503.4   $&$ 466.1   $&$ 503.9$&$    489.0   $&$ 457.4   $&
$   386.8   $\\
\hline\\[0.5ex]
\multicolumn{5}{l}{\TBADEX}
\\[0.6ex]
\multicolumn{1}{r|}{$A_0$}  
    &$  519.3   $
    &$  240.8   $&$ 151.1   $&$ 68.6$&$ 59.0    $&$ 56.9    $
    &$  502.9   $&$ 463.3   $&$ 500.3$&$    484.7   $&$ 453.0   $
    &$  355.8   $
    \\
\multicolumn{1}{r|}{$A_{max}$}  &$  546.8   $&
$   247.8   $&$ 155.2   $&$ 69.9$&$ 59.1    $&$ 57.1    $&
$   525.5   $&$ 486.2   $&$ 524.0$&$    510.0   $&$ 476.9   $&
$   366.1   $\\
\multicolumn{1}{r|}{$A_{mean}$} &$  528.5   $&
$   243.1   $&$ 152.2   $&$ 68.9$&$ 58.9    $&$ 56.5    $&
$   510.6   $&$ 471.2   $&$ 508.3$&$    493.2   $&$ 461.0   $&
$   359.7   $\\
\hline\\[0.5ex]

\multicolumn{8}{l}{\bf FSI (steady flow)}
\\[0.8ex]
\multicolumn{1}{r|}{\TBADOR}    
    &$  522.1   $
    &$  241.6   $&$ 151.6   $&$ 68.9$&$ 59.1    $&$ 57.0    $
    &$  505.2   $&$ 465.4   $&$ 502.6$&$    487.0   $&$ 455.4   $
    &$  385.2   $\\
\hline\\[0.5ex]
    
\multicolumn{8}{l}{\bf STL model (``flow-off'')}
\\[0.8ex]
\multicolumn{1}{c|}{\TBADOR }
&$  492.6   $
&$  235.6   $&$ 149.2   $&$ 67.6$&$ 58.4    $&$ 56.6    $
&$  476.3   $&$ 436.3   $&$ 475.0$&$    458.7   $&$ 429.3   $
&$  365.5   $\\\hline
\end{tabular}
\end{center}
\caption*{\textbf{S9. Cross-sectional area} (in \si{mm^2}) at twelve landmarks  for pulsatile flow, steady flow, and ``flow-off'' modes.
$A_{0}$ (area at first frame of the cardiac cycle) and $A_{max}$ (maximum area of cardiac cycle) were considered to define end-diastolic and peak-systolic cross-sectional area, respectively.
$A_{mean}$ reports the area averaged over the cardiac cycle. 
For landmark label definition see Fig. 2a of main article.
Refer to Supplementary Fig. S10 for data plots.
} 
\label{Table:Area}
\end{table}

\newpage
\begin{figure}[ht]
    \centering
    \includegraphics[]{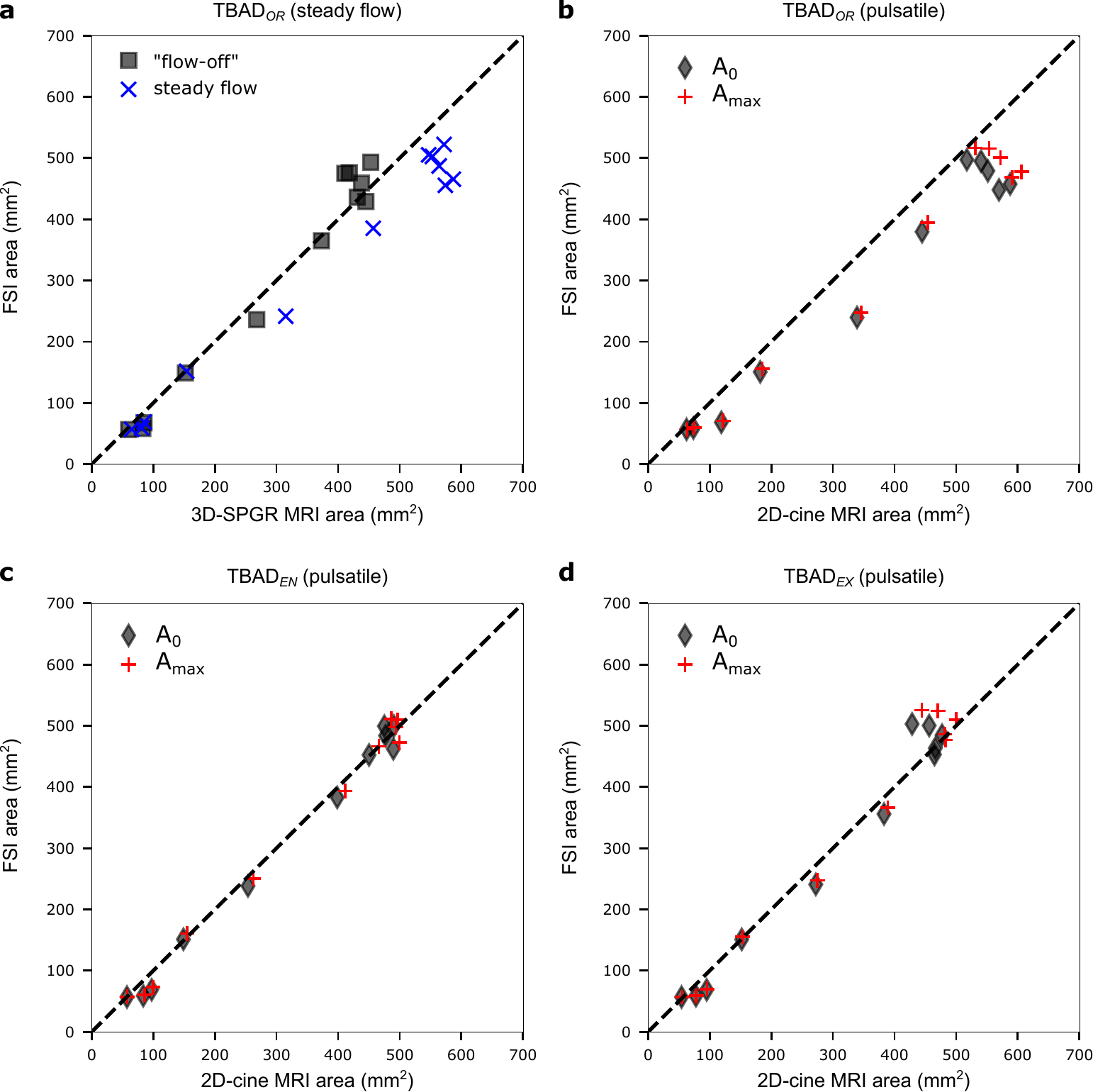}
    \caption*{
        \textbf{S10. Absolute cross-sectional area} evaluated at twelve landmarks with 3D-SPGR or 2D-cine MRI of the 3D-printed model (horizontal axis) and based on the deformable structural domain, i.e. aortic wall, in FSI simulations (vertical axis):
        (a) ``flow-off'' and steady flow measurements; 
        (b, c, d) first frame ($A_0$, end-diastolic) and maximum ($A_{max}$, peak-systolic) area measurements in pulsatile mode for each model. Refer to Supplementary Table S9 for underlying data.}
    \label{FIG_areaCorrelationPlots_v2}
\end{figure}

\clearpage
\noindent \textbf{Supplementary video file description}: In all animated vector visualizations, cycle length was stretched from \SI{1}{\second} to \SI{2}{\second} to better display complex patterns, and the video file was exported with 24 fps.

\vspace{0.5cm}
\noindent\textbf{S4} 4D-flow MRI velocity vector visualizations in three TBAD models. Cycle length was slowed down to \SI{2}{\second}.

\vspace{0.5cm}
\noindent\textbf{S5} CFD-FSI simulations displaying velocity vectors of fluid domain in three TBAD models.

\vspace{0.5cm}
\noindent\textbf{S6} Entry tear close-up view of velocity vector visualizations of 4D-flow MRI and CFD-FSI simulations.

\vspace{0.5cm}
\noindent\textbf{S7} Entry tear close-up view of velocity vector visualizations of 4D-flow MRI and CFD-FSI simulations. Identical data as in video S6, but with camera view rotated around aorta long axis.

\vspace{0.5cm}
\noindent\textbf{S8} Exit tear close-up view of velocity vector visualizations of 4D-flow MRI and CFD-FSI simulations.

\end{document}